\documentclass[10pt]{article}
\usepackage[square,authoryear]{natbib}
\usepackage{graphicx}
\usepackage{marsden_article}

\usepackage{epstopdf,framed}
\usepackage{pdfsync}
\begin{document}
\newtheorem{remark}[theorem]{Remark}
\newtheorem{example}[theorem]{Example}

\title{A Variational Formulation of Nonequilibrium Thermodynamics for Discrete Open Systems with Mass and Heat Transfer}
\vspace{-0.2in}

\newcommand{\todoFGB}[1]{\vspace{5 mm}\par \noindent
\framebox{\begin{minipage}[c]{0.95 \textwidth} \color{red}FGB: \tt #1
\end{minipage}}\vspace{5 mm}\par}


\author{\hspace{-1cm}
\begin{tabular}{cc}
Fran\c{c}ois Gay-Balmaz &
Hiroaki Yoshimura
\\  Laboratoire de m\'et\'eorologie dynamique  & School of Science and Engineering
\\ CNRS \& Ecole Normale Sup\'erieure & Waseda University
\\  24 Rue Lhomond 75005 Paris, France & Okubo, Shinjuku, Tokyo 169-8555, Japan \\ francois.gay-balmaz@lmd.ens.fr & yoshimura@waseda.jp\\
\end{tabular}\\\\
}

\date{}
\maketitle
\vspace{-0.3in}

\begin{center}
\abstract{We propose a variational formulation for the nonequilibrium thermodynamics of discrete open systems, i.e., discrete systems which can exchange mass and heat with the exterior. Our approach is based on a general variational formulation for systems with time-dependent nonlinear nonholonomic constraints and time-dependent Lagrangian. For discrete open systems, the~time-dependent nonlinear constraint is associated with the rate of internal entropy production of the system. We show that this constraint on the solution curve systematically yields a constraint on the variations to be used in the action functional. The proposed variational formulation is intrinsic and provides the same structure for a wide class of discrete open systems. We illustrate our theory by presenting examples of open systems experiencing mechanical interactions, as well as internal diffusion, internal heat transfer, and their cross-effects. Our approach yields a systematic way to derive the complete evolution equations for the open systems, including the expression of the internal entropy production of the system, independently on its complexity. It might be especially useful for the study of the nonequilibrium thermodynamics of biophysical systems.}
\vspace{2mm}

\end{center}
\tableofcontents

\section{Introduction}

The goal of this paper is to present a variational formulation for the nonequilibrium thermodynamics of open discrete (finite dimensional) systems, i.e., systems which may exchange heat as well as matter with their surrounding. Such systems, out of equilibrium, appear naturally in engineering, physics, biology, chemistry, social science, the archetypal example being the case of the living cell. A first step in the study of the nonequilibrium thermodynamics of a given system is the determination of the complete set of evolution equations, written as a system of ordinary differential equations for discrete systems, which allows to determine the values of all the variables of the system, at all times. Such a step is nontrivial, even for closed systems, especially in the case when several irreversible processes of various nature interact with each other. A celebrated example being the case of the adiabatic piston, whose final configuration can only be determined by solving the ODEs of evolution, see~\cite{Gr1999}.

In our previous works~\cite{GBYo2017a, GBYo2017b}, we developed a Lagrangian variational formulation for the nonequilibrium thermodynamics of closed discrete systems by following Stueckelberg's phenomenological approach, in which we extended the Lagrange-d'Alembert principle for nonlinear nonholonomic mechanics to encode the internal entropy production due to irreversible processes into a class of nonlinear constraints of thermodynamic type. Such a variational approach is based on two types of constraints, a {phenomenological constraint} imposed on the solution curve and a {variational constraint} imposed on the variations to be considered. As shown in~\cite{GBYo2017a, GBYo2017b}, these two constraints are related to each other in a very systematic way.  In addition, our approach gives a form of the entropy production equation that is extremely well-suited for the choice of the phenomenological relation between the thermodynamic fluxes and the thermodynamic forces, in accordance with the second law of thermodynamics. To obtain such a systematic variational framework, the notion of {thermodynamical displacements} is introduced, which enables us to systematically obtain the variational constraint from the phenomenological constraint. As shown in~\cite{GBYo2017a, GBYo2017b}, the stationarity condition for a certain action functional involving the total Lagrangian of the system yields the complete evolution equations for all the variables of the system. The variational formulation in~\cite{GBYo2017a, GBYo2017b} is illustrated with several examples of discrete systems such as mechanical systems with friction, electric circuits with resistor, chemical~reactions, diffusions across membranes, etc.

In this paper, we extend this variational formulation of nonequilibrium thermodynamics to the case of  discrete {open} systems. Passing from closed to open systems introduces an important change in the constraints to be used, namely, the constraints generically become explicitly {time-dependent}, due to the power transfer associated to matter and heat exchange with the exterior through the ports. More~precisely, we have to extend our phenomenological constraint to a {kinematic} constraint by taking into account of the time-dependent boundary conditions associated to exchanges with the exterior for the derivation of the entropy production. To reach this goal, we first propose a novel theory of Lagrangian variational formulation for systems with time-dependent  nonlinear nonholonomic constraints, in which the kinematic and variational constraints are related in a very specific way. When~applying this variational formulation to open thermodynamic system, the constraint is given by the equation for the rate of entropy production of the system. Our approach can handle systems with several ports, experiencing mechanical interactions, internal diffusion, internal heat transfer, and their~cross-effects.

The variational approach presented in this paper enables an effective derivation of the complete evolution equations for all the variables of the open system. It thus represents an essential modeling tool for the dynamics of open systems, in the same way as the Hamilton principle in Lagrangian mechanics (see, e.g.,~\cite{GBPu2015}, for a recent use of the Hamilton variational principle as a modeling tool for fluid-structure interaction) and its Lagrange-d'Alembert version for nonholonomic mechanical systems. A preliminary illustration of this variational feature is considered in Section \ref{subsec_mechanics}, where~we propose a model for an open system which takes into account of the exchange of mechanical momentum in the dynamics and in the entropy production equation. We project to extensively exploit our formulation as a modeling tool for the nonequilibrium thermodynamics of open systems in the future.
Besides its modeling purpose, our approach is also useful for the derivation of structure preserving numerical schemes for nonequilibrium thermodynamics, based on discrete versions of the variational formulation, in a similar way with the case of variational integrators in classical mechanics \cite{MaWe2001}. We made initial steps in this direction in~\cite{GBYo2018b}. The description of the nonequilibrium thermodynamics used in this paper assumes the local thermodynamic equilibrium condition.

\color{black}

Our paper is organized as follows. In Section~\ref{Section_2}, we review the fundamental ingredients for the nonequilibrium thermodynamic of discrete open systems by following Stueckelberg's formulation of the two laws. In Section~\ref{Section_3}, we first introduce an abstract variational formulation for systems with nonlinear time-dependent nonholonomic constraints. Then, we present the simplest case of an open system with a single chemical component in a single compartment with matter transfer through one entry port. Further, we explore the case of a discrete open system with a single chemical component experiencing internal diffusion through several compartments, as well as matter transfer through multi-entry ports, without heat transfer. We illustrate our theory with some examples such as an open system of flowing fluid carrying kinetic energy into a piston device, an open system with a single chemical component experiencing diffusion between two compartments and transfer into or out of the system through two ports, as well as open systems  with diffusion of a single chemical species through a composite membrane made of both serial and parallel arrays of several elements. In Section~\ref{sec_heat_matter}, we~investigate the more general class of discrete open systems, having several entropy variables, and hence several temperatures. For instance, we consider a single chemical component in several compartments, experiencing internal diffusion, heat transfer, the associated cross-effects, as well as matter exchange through multi-entry ports, and external heating. We illustrate our variational formulation with several examples such as an open system experiencing diffusion and heat conduction between two compartments, and an open system with heat conduction and diffusion of a single chemical species through a composite membrane made of a parallel array of several elements. In~Section~\ref{Section_5}, some concluding remarks are given as well as future directions.

\section{Simple Discrete Open Systems}\label{Section_2}

In this paper, we shall follow the phenomenological approach to nonequilibrium thermodynamics developed by Stueckelberg~\cite{StSc1974}, which provides a macroscopic dynamic theory for accounting for irreversible processes. Such an approach enables the treatment of nonequilibrium thermodynamics as a natural extension of classical mechanics.
Before going into details, we 
make a brief review on the fundamental notions that we employ by following Stueckelberg's approach (see, for~instance,~\cite{Gr1999,FeGr2010,GrBr2011}), where two state functions, the {energy} and the {entropy}, are used to obey the two fundamental laws of thermodynamics, which are formulated as first order differential equations.


Let us denote by $ \boldsymbol{\Sigma}  $ a physical system and by $ \boldsymbol{\Sigma} ^{\rm ext}$ its exterior. The state of the system is defined by a set of mechanical variables and a set of thermal variables. State functions are functions of these variables. Stueckelberg's formulation of the two laws is given as follows.

\medskip

\noindent {\bf First law:} For every system $ \boldsymbol{\Sigma} $, there exists an extensive scalar state function $E$, called {energy}, which~satisfies
\[
\frac{d}{dt} E(t) = P^{\rm ext}_W(t)+P^{\rm ext}_H(t)+P^{\rm ext}_M(t),
\]
where $t$ denotes {time}, $ P^{\rm ext}_W$ is the {{power associated to the work done on the system}}, $P^{\rm ext}_H$ is the {power associated to the transfer of heat into the system}, and $P^{\rm ext}_M$ is the {power associated to the transfer of matter into the system} (As we will recall below, to matter transfer is associated a transfer of work and heat. By convention, $ P^{\rm ext}_W$ and $P^{\rm ext}_H$ denote uniquely the power associated to work and heat that is {not associated to a transfer of matter.}).

Given a {thermodynamic system}, the following terminology is generally adopted:
\begin{itemize}
\item A system is said to be {closed} if there is no exchange of matter, i.e.,  $P^{\rm ext}_M(t)=0$. When $P^{\rm ext}_M(t) \ne 0$ the system is said to be {open}.
\item 
A system is said to be {adiabatically closed} if it is closed and there is no heat exchanges, i.e., $P^{\rm ext}_M(t)=P^{\rm ext}_H(t)=0$. 
\item 
A system is said to be {isolated} if it is adiabatically closed and there is no mechanical power exchange, i.e., $P^{\rm ext}_M(t)=P^{\rm ext}_H(t)=P^{\rm ext}_W(t)=0$.
\end{itemize}

From the first law, it follows that the {energy of an isolated system is constant}.
\medskip

\noindent {\bf Second law:} For every system $ \boldsymbol{\Sigma} $, there exists an extensive scalar state function $S$, called {entropy}, which obeys the following two conditions (see~\cite{StSc1974}, p. 23)

\begin{itemize}
\item[(a)] Evolution part:\\
If the system is adiabatically closed, the entropy $S$ is a non-decreasing function with respect to time, i.e., 
\[
\frac{d}{dt} S(t)=I(t)\geq 0,
\]
where $I(t)$ is the {entropy production rate} of the system accounting for the irreversibility of internal~processes.
\item[(b)] Equilibrium part:\\
If the system is isolated, as time tends to infinity the entropy tends towards a finite local maximum of the function $S$ over all the thermodynamic states $ \rho $ compatible with the system, i.e., 
\[
\lim_{t \rightarrow +\infty}S(t)= \max_{ \rho \; \text{compatible}}S[\rho ].
\]
\end{itemize}

By definition, the evolution of an isolated system is said to be {reversible} if $I(t)=0$, namely, the entropy is constant. In general, the evolution of a system $ \boldsymbol{\Sigma} $ is said to be {reversible}, if the evolution of the total isolated system with which $ \boldsymbol{\Sigma} $ interacts is reversible.
\medskip

\noindent {\bf Expression of the first law for open systems.} Consider an open system with several ports, $a=1,...,A$, through which matter can flow into or out of the system. We suppose, for simplicity, that the system involves only one chemical species and denote by $N$ the number of moles of this species. The mole balance equation is
\[
\frac{d}{dt}N=\sum_{a=1}^A \mathcal{J}^a,
\]
where $\mathcal{J}^a$ is the molar flow rate \textit{into} the system through the $a$-th port, so that $\mathcal{J}^a>0$ for flow into the system and $\mathcal{J}^a<0$ for flow out of the system.

As matter enters or leaves the system, it carries its internal, potential, and kinetic energy. This~energy flow rate at the $a$-th port is the product $\mathsf{E}^a\mathcal{J}^a$ of the energy per mole (or molar energy) $\mathsf{E}^a$ and the molar flow rate $\mathcal{J}^a$ at the $a$-th port. In addition, as matter enters or leaves the system it also exerts work on the system that is associated with pushing the species into or out of the system. The~associated energy flow rate is given at the $a$-th port by $p^a\mathsf{V}^a\mathcal{J}^a$, where $p^a$ and $\mathsf{V}^a$ are the pressure and the molar volume of the substance flowing through the $a$-th port.

From this we can write the first law of thermodynamics as
\begin{equation}\label{law1_explicit}
\frac{d}{dt} E  =P^{\rm ext}_W+P^{\rm ext}_H+P^{\rm ext}_M, \quad\text{with}\quad P^{\rm ext}_M= \sum_{a=1}^A\mathcal{J}^a(\mathsf{E}^a+ p^a\mathsf{V}^a).
\end{equation}

We refer, for instance, to~\cite{Sa2006,KlNe2011} for the detailed explanations of the first law for open systems. We~recall below the definition of simple systems, which plays a main role in our variational formulation.

\medskip

\noindent {\bf Discrete simple systems.} A {discrete system} $ \boldsymbol{\Sigma}$ is a collection $ \boldsymbol{\Sigma} =\cup_{A=1}^N \boldsymbol{\Sigma} _A $ of a finite number of interacting simple systems $ \boldsymbol{\Sigma} _A $. By definition, following~\cite{StSc1974}, a {simple} system $ \boldsymbol{\Sigma}$ is a macroscopic system for which one (scalar) thermal variable and a finite set of mechanical variables are sufficient to describe entirely the state of the system. From the second law of thermodynamics, we can always choose the thermal variable as the entropy $S$.

\section{Variational Formulation of Discrete Open Simple Systems}\label{Section_3}

In this section, we develop a variational formulation for discrete open thermodynamic systems that are assumed to be {simple}. As recalled above, such systems are described by a single entropy variable and, therefore, a uniform temperature can be attributed to the system. The case of non-simple systems will be treated in Section \ref{sec_heat_matter}.

We start in Section \ref{subsec_abstract} by introducing a general variational approach for the treatment of time-dependent nonholonomic systems with nonlinear constraints. This abstract variational formulation extends to the time dependent case the one developed in~\cite{GBYo2017a} for closed thermodynamic systems and provides the general framework for the variational formulation for open systems. In~Section~\ref{sec_foundations} we develop the variational formulation for open simple systems in a particular setting, before extending it to include mechanical interactions in Section \ref{subsec_mechanics} and internal diffusion in Section \ref{subsec_diffusion}.
Several examples are treated in Section \ref{subsec_examples}, while Section \ref{link_abstract} tightens the link between the variational formulation for open systems and the abstract formulation presented in Section \ref{subsec_abstract}.

\subsection{A Variational Formulation for Time-Dependent Nonlinear Nonholonomic Systems}\label{subsec_abstract}

In this section, we present an abstract variational framework for the treatment of open thermodynamic systems later. The variational formulation consists in computing the critical curve of an action functional by imposing two kinds of constraints: a {variational constraint} which imposes constraints on the variations of the curve, and a {kinematic constraint} which imposes constraints on the critical curves. These two constraints are related in a very specific way. This setting extends the variational formulation proposed in~\cite{GBYo2017a,GBYo2017b,GBYo2018a} for closed thermodynamic systems. In the particular case of linear constraints, the variational formulation recovers the Lagrange-d'Alembert principle for nonholonomic mechanics, see, e.g.,~\cite{Bl2003}. A general variational setting for mechanics with nonlinear constraints is described in~\cite{CeIbdLdD2004,Ma1998}. For the variational formulation of time-dependent nonholonomic mechanics with {affine constraints}, see~\cite{VaYoLe2012}.

Let $Q$ be a configuration manifold with local coordinates $q=(q^{1},...,q^{n})$ and let $\mathbb{R}\times Q$ be the extended configuration manifold which includes the time variable $t \in\mathbb{R}$. We consider the vector bundle $\mathbb{R} \times TQ$ over $\mathbb{R}\times Q$ with local coordinates $(t, q,v)$ and the vector bundle $(\mathbb{R} \times TQ) \times_{\mathbb{R} \times Q} (\mathbb{R} \times TQ)$ over $\mathbb{R}\times Q$ with local coordinates $(t, q,v, \delta q)$.

Consider a time-dependent constraint $C_{V} \subset (\mathbb{R} \times TQ) \times_{\mathbb{R} \times Q} (\mathbb{R} \times TQ)$ of the form\vspace{-6pt}
\begin{align*}
C_{V}&=\{(t,q,v,\delta{q}) \in (\mathbb{R} \times TQ) \times_{\mathbb{R} \times Q} (\mathbb{R} \times TQ) \mid A_{i}^{r}(t,q,v)\delta{q}^{i} +B^{r}(t,q,v)=0,\\
&\hspace{10cm} r=1,...,m<n \}.
\end{align*}
Such a constraint defines for each $(t,q,v) \in \mathbb{R} \times TQ$ an affine subspace $C_{V}(t,q,v)\subset T_qQ$. By~considering, for each $(t,q,v)$, the linear subspace $\widehat{C}_{V}(t,q,v)$ associated to $C_{V}(t,q,v)$, we get the time-dependent variational constraint
\begin{equation*}\label{hatCV}
\widehat{C}_{V}=\left\{(t,q,v,\delta{q}) \in (\mathbb{R} \times TQ) \times_{\mathbb{R} \times Q} (\mathbb{R} \times TQ) \mid A_{i}^{r}(t,q,v)\delta{q}^{i} =0,\; r=1,...,m<n\right\}.
\end{equation*}
Given $C_V$, we define the associated nonlinear nonholonomic kinematic constraint $C_{K}\subset \mathbb{R} \times TQ$ by
\begin{equation*}\label{CK}
\begin{split}
C_{K}&=\left\{ (t,q,v)\in \mathbb{R} \times TQ \mid v \in  C_{V}(t,q,v)\right\} \\
&= \left\{(t,q,v)\in \mathbb{R} \times TQ \mid A_{i}^{r}(t,q,v)v^{i} +B^{r}(t,q,v)=0,\; r=1,...,m<n\right \}.
\end{split}
\end{equation*}
We note that the function $A_i^r$ and $B^r$ may depend explicitly on the time $t$.

Let $L$ be a time-dependent Lagrangian defined on $\mathbb{R} \times TQ$ and let $F^{\rm ext}: \mathbb{R} \times TQ \to T^{\ast}Q$ be an external force field assumed to be fiber preserving, i.e., $F^{\rm ext}(t,q,v)\in T^*_qQ$.
We consider the following {Lagrange-d'Alembert principle with time-dependent nonlinear constraints and external force field}:
\begin{equation}\label{Vcond_abstract}
\delta \int_{t_{1}}^{t_{2}}  L(t,q,\dot{q}) dt+\int_{t _1 }^{t _2 } \left\langle F^{\rm ext}(t, q, \dot q), \delta q \right\rangle dt=0,\qquad \textsc{Variational Condition}
\end{equation}
with respect to variations $\delta q(t) \in \hat{C}_{V}(t, q, \dot{q})$, i.e., 
\begin{equation}\label{VC_abstract}
A_{i}^{r}(t,q,\dot{q})\delta{q}^{i} =0, \qquad \textsc{Variational Constraint}
\end{equation}
with $\delta{q}(t_{1})=\delta{q}(t_{2})=0$, and where the curve $q(t)$ satisfies the the nonlinear constraint $C_{K}$, i.e,
\begin{equation}\label{PC_abstract}
A_{i}^{r}(t,q,\dot{q})\dot{q}^{i} +B^{r}(t,q,\dot{q})=0 ,\qquad \textsc{Kinematic Constraint}.
\end{equation}

By a direct computation, using Lagrange multipliers $\lambda_{r},\; r=1,...,m$, we get that a curve $q(t)$ is critical for the variational formulations \eqref{Vcond_abstract}--\eqref{PC_abstract} if and only if it is a solution of the {time-dependent nonlinear nonholonomic Lagrange-d'Alembert equations with external forces}:
\vspace{12pt}
\begin{equation}\label{evo_eqn_nh1}
\left\{
\begin{array}{l}
\vspace{0.2cm}\displaystyle\frac{d}{dt}\frac{\partial L}{\partial \dot{q}^{i}}(t,q,\dot{q})- \frac{\partial L}{\partial q^{i}}(t,q,\dot{q})=\lambda_{r}A_{i}^{r}(t,q,\dot{q})+F^{\rm ext}_{i}(t, q, \dot q),\\
\displaystyle A_{i}^{r}(t,q,\dot{q})\dot{q}^{i} +B^{r}(t,q,\dot{q})=0.
\end{array}\right.
\end{equation}

Associated to the Lagrangian $L(t, q,\dot{q})$, we define the energy $E(t, q,\dot{q})$ on $\mathbb{R} \times TQ$ as
\[
E(t,q,\dot{q})=\frac{\partial L}{\partial \dot{q}^{i}}\dot{q}^{i}-L(t,q,\dot{q}).
\]
Along a solution of the evolution Equation \eqref{evo_eqn_nh1}, we get the {energy balance equation}
\begin{equation}\label{energy_t}
\frac{d}{dt}E(t,q,\dot{q})=F^{\rm ext}_{i}(t, q, \dot q)\dot{q}^{i}-\lambda_{r}B^r-\frac{\partial L}{\partial t}.
\end{equation}

As we shall show later, the variational formulation for discrete open thermodynamic systems falls into the abstract variational setting  \eqref{Vcond_abstract}--\eqref{PC_abstract}. In this case, the system \eqref{evo_eqn_nh1} yields the complete evolution equations for the thermodynamic system, in which the nonlinear constraint gives the entropy equation, in agreement with the second law of thermodynamics. The energy balance Equation \eqref{energy_t} yields the first law of thermodynamics.

\subsection{Foundations of the Variational Formulation for Discrete Open Thermodynamic Systems}\label{sec_foundations}

In order to introduce our variational formulation, we shall first consider a particular case of simple discrete system, namely, the case of a system with a single chemical component $N$ in a single compartment with constant volume $V$. Recall that, since the system is assumed to be simple, there is a single entropy $S$ attributed to the whole system. Non-simple systems will be considered in Section~\ref{sec_heat_matter}.
We assume that there is no heat and work exchanges, except the ones associated to the transfer of matter. 
We also ignore all the mechanical effects, which will be included later in Section~\ref{subsec_mechanics}. In this particular situation, the energy of the system is given by the internal energy written as $U=U(S,N)$, since $V=V_0$ is constant. The balance of mole and the balance energy, i.e., the first law, are respectively given by
\[
\frac{d}{dt}N=\sum_{a=1}^A \mathcal{J}^a, \quad \frac{d}{dt} U  = \sum_{a=1}^A\mathcal{J}^a(\mathsf{U}^a+ p^a\mathsf{V}^a)=\sum_{a=1}^A\mathcal{J}^a\mathsf{H}^a,
\]
see \eqref{law1_explicit}, where $\mathsf{H}^a=\mathsf{U}^a+ p^a\mathsf{V}^a$ is the molar enthalpy at the $a$-th port and where $\mathsf{U}^a$, $ p^a$, and $\mathsf{V}^a$ are respectively the molar internal energy, the pressure and the molar volume at the $a$-th port. From these equations and the second law, one obtains the equations for the rate of change of the entropy of the system as
\begin{equation}\label{S_dot_simple}
\frac{d}{dt}S= I+\sum_{a=1}^A \mathsf{S}^a\mathcal{J}^a,
\end{equation}
where $\mathsf{S}^a$ is the molar entropy at the $a$-th port and $I$ is the rate of internal entropy production of the system given by
\begin{equation}\label{I_simple_example}
I= \frac{1}{T}\sum_{a=1}^A \mathcal{J}^a\left(\mathsf{H}^a-T\mathsf{S}^a- \mu \right),
\end{equation}
with $T= \frac{\partial U}{\partial S}$ the temperature and $\mu=\frac{\partial U}{\partial N}$ the chemical potential.
For our variational treatment, it is useful to rewrite the rate of internal entropy production as\vspace{-6pt}
\[
I= \frac{1}{T}\sum_{a=1}^A \left[\mathcal{J}^a_S(T^a-T)+\mathcal{J}^a(\mu^a- \mu )\right],
\]
where we defined the entropy flow rate $\mathcal{J}^a_S:=\mathsf{S}^a\mathcal{J}^a$ and also used the relation $\mathsf{H}^a=\mathsf{U}^a+ p^a\mathsf{V}^a= \mu^a +T^a\mathsf{S}^a$. The thermodynamic quantities known at the ports are usually the pressure and the temperature $p^a$, $T^a$, from which the other thermodynamic quantities, such as $\mu^a=\mu^a(p^a,T^a)$ or $\mathsf{S}^a=\mathsf{S}^a(p^a,T^a)$ are deduced from the state equations of the gas.

\begin{remark}{\rm 
There are several representations for the rate of internal entropy production given in \eqref{I_simple_example}, namely, we can rewrite $I$ as
\begin{equation*}
\begin{split}
I= \frac{1}{T}\sum_{a=1}^A \mathcal{J}^a\left[T(\mathsf{S}- \mathsf{S}^{a})-(\mathsf{U}-\mathsf{U}^{a})-(p\mathsf{V}-p^{a}\mathsf{V}^{a}) \right]
\end{split}
\end{equation*}
as well as
\begin{equation*}
\begin{split}
I= \frac{1}{T}\sum_{a=1}^A \mathcal{J}^a\left[(\mathsf{H}^a-\mathsf{H})-T(\mathsf{S}^a- \mathsf{S}) \right].
\end{split}
\end{equation*}}
\end{remark}

\noindent{\bf Variational formulation.} 
{Let $W$, $\Gamma$, and $\Sigma$ be the molar and thermal displacements and the internal entropy of the system respectively, each of whose interpretation will be explained later.}
Given the molar flow rates $\mathcal{J}^a$, the entropy flow rates $\mathcal{J}_S^a$, the temperatures $T^a$, and the chemical potentials $\mu ^a$ at the ports $a=1,...,A$, we consider the following variational formulation:\vspace{-6pt}
\begin{equation}\label{VCond_simple}
\delta \int_{t _1 }^{ t _2} \Big[ -U(S, N)+\dot W N +(S- \Sigma ) \dot \Gamma \Big] dt =0 , \quad\;\;\; \textsc{Variational Condition}
\end{equation}
where the curves satisfy the 
nonlinear nonholonomic constraint \vspace{-6pt}
\begin{equation}\label{PC_simple}
-\frac{\partial U}{\partial S}\dot \Sigma   =\sum_{a=1}^A \Big[\mathcal{J} ^a\big( \dot W- \mu ^a\big) + \mathcal{J} ^a _S\big( \dot \Gamma - T ^a\big) \Big],\quad \textsc{{Kinematic} Constraint}
\end{equation}
and with respect to variations subject to the constraint\vspace{-6pt}
\begin{equation}\label{VCons_simple}
-\frac{\partial U}{\partial S}\delta  \Sigma   =  \sum_{a=1}^A \Big[ \mathcal{J} ^a\delta  W  + \mathcal{J} ^a_S\delta  \Gamma  \Big],\quad \textsc{Variational Constraint}
\end{equation}
with $ \delta W(t_1)=\delta W(t_2)=0$.
\medskip

\begin{remark}{\rm
We call the internal entropy production \eqref{PC_simple} the {kinematic constraint, where the internal entropy production, in general, is described not only by the phenomenological relations obtained experimentally but also by the pressures, temperatures and the molar flow rates at the external ports. This equation is generally a nonlinear nonholonomic constraint imposed on the curve of the state variables in the variational formulation, see \eqref{PC_abstract}.} We will later explain in details how the variational {formulations}~\eqref{VCond_simple}--\eqref{VCons_simple} arises as a special instance of the abstract variational formulations \eqref{Vcond_abstract}--\eqref{PC_abstract}.}
\end{remark}

\begin{remark}{\rm
In the variational formulations given in \eqref{VCond_simple}--\eqref{VCons_simple}, we employed the new variables, $ \Gamma$ and $W$, such~that  $\dot{\Gamma}=T$ and $\dot{W}=\mu$, where we interpreted the variable $\Gamma$ as the {thermal displacement} and the variable $W$ as the {molar displacement}. These are examples of {thermodynamic displacements}, as introduced in~\cite{GBYo2017a,GBYo2017b}, whose rate of change coincide with the affinity of the irreversible process. In particular, note~that $ \Gamma$ is a monotonically increasing real function of time $t$ since the temperature $T$ takes positive real values. Note also that the notion of thermal displacement was first used by~\cite{He1884} and in the continuum setting by~\cite{GrNa1991}. Refer to the Appendix of~\cite{Po2009} for an historical account. The variables $ \Gamma$ and  $W$ are  conjugate variables with $S$ (or $ \Sigma$) and $N$, respectively. Moreover, we used another new variable $\Sigma$, which is not the same object as the entropy of the system $S$ but it denotes the {internal entropy of the system}. In fact, $\dot{\Sigma}$ indicates the {internal} entropy production, with $\dot \Sigma \geq 0$, sometimes denoted as $d_iS/dt$ in the standard literature (see, for instance,  Equation  \eqref{I_simple_example} on page 21 in~\cite{deGrootMazur1969}), while the total rate of entropy production of the system, $\dot S$, does not need to be positive if the system is not adiabatically closed.}
\end{remark}

\noindent{\bf Comments and structure of the variational formulation.}
Before applying the variational formulation to derive the equations, we first comment on the variables and functions involved as well as on the structure of this formulation. The variational formulations \eqref{VCond_simple}--\eqref{VCons_simple} gives conditions on the curve $c(t)=\big(\Gamma(t), W(t), N(t),S(t),\Sigma(t)\big)$ that ultimately gives a system of ODEs for this curve. This curve involves the thermodynamic quantities associated to the system, namely, the number of moles $N(t)$, the total entropy $S(t)$ and the internal entropy $\Sigma(t)$ of the system, as well as the thermodynamic displacements $\Gamma(t)$, $W(t)$ whose interpretation will be explained below. Recall that the knowledge of a fundamental state equation of the system, such as $U=U(S,V,N)$, allows one to deduce all the other thermodynamic relations.

The thermodynamic quantities associated to the ports $a=1,...,A$, are denoted with an exponent~$a$, such as $T^a$ and $\mu^a$. They satisfy the same relations as the corresponding quantities of the system, because they represent the same species, i.e., are obtained from the same state equation.
Given a port~$a$, if matter is flowing into the system through that port from the exterior, then we will assume that the thermodynamic quantities are known from the experimenter, i.e., they are prescribed functions of time such as $T^a=T^a(t)$ and $\mu^a=\mu^a(t)$, which characterise the thermodynamic properties of the matter flowing into the system. On the other hand, if matter is flowing out of the system through that port, then the associated thermodynamic quantities coincide with the thermodynamic quantities of the system, such as, $T^a=T$, $\mu^a=\mu$. In this case, these quantities are not assumed to be known, they are computed via the evolution equations. If matter is flowing in, we shall assume that the molar flow rate $\mathcal{J}^a>0$ is a prescribed function of time, i.e., $\mathcal{J}^a=\mathcal{J}^a(t)$. If matter is flowing out of the system then $\mathcal{J}^a<0$ is usually computed from the other variables and the physical properties of the system.
In the system considered here, since there is no additional heating from the exterior, the entropy flow rate $\mathcal{J}^a_S$ is given in terms of the molar flow rate $\mathcal{J}^a$ as $\mathcal{J}^a_S=\mathsf{S}^a\mathcal{J}^a$, where $\mathsf{S}^a$ is the molar entropy at the $a$-th port.

We note that the variational constraint \eqref{VCons_simple} follows from the phenomenological constraint \eqref{PC_simple} by formally replacing the rates by the corresponding virtual displacements, i.e., $\dot{x} \rightarrow \delta x$, and by removing all the terms that depend uniquely on the ports $a=1,...,A$, i.e., the terms $\mu ^a\mathcal{J} ^a$ and $T ^a \mathcal{J} ^a_S$. Such a systematic correspondence between the phenomenological and variational constraints, was already verified for the variational formulation of closed systems, see~\cite{GBYo2017a,GBYo2017b}. It is extended here to the case of open systems. Note that the action functional in \eqref{VCond_simple} has the same form as that in the case of closed systems, see~\cite{GBYo2017a}. As~will be shown later, this systematic relation between the two constraints still holds in the more general cases of discrete open systems considered below.

\medskip

\noindent{\bf Application of the variational formulation.}
By taking the variation of the integral in \eqref{VCond_simple} and using the variational constraint \eqref{VCons_simple}, we get the following {evolution equations}:\vspace{-6pt}
\begin{equation}\label{resulting_conditions_simple}
\begin{aligned}
\delta S:&\quad\dot \Gamma = \frac{\partial U}{\partial S}, \\
\delta W:&\quad\dot N= \sum_{a=1}^A\mathcal{J} ^a, \\
\delta N:&\quad\dot W=  \frac{\partial U}{\partial N}, \\
\delta \Gamma :&\quad\dot S=\dot \Sigma +\sum_{a=1}^A\mathcal{J}^a_S.
\end{aligned}
\end{equation}
From the first and third equations in \eqref{resulting_conditions_simple}, temperature is obtained as the rate of the thermal displacement $\Gamma$ while the chemical potential as the rate of the molar displacement $W$, since the temperature and the chemical potential of the system are defined by $T:=\frac{\partial U}{\partial S}$ and $\mu:=\frac{\partial U}{\partial N}$.

By inserting the conditions \eqref{resulting_conditions_simple} into the phenomenological constraint \eqref{PC_simple}, we get the desired entropy equation in terms of $T$ and $\mu$, namely
\[
T\dot S=\sum_{a=1}^A \left[\mathcal{J}^a(\mu^a-\mu) +\mathcal{J}_S^a(T^a-T)\right]+ T\sum_{a=1}^A \mathcal{J} ^a_S= \sum_{a=1}^A \mathcal{J} ^a( \mathsf{H} ^a- T\mathsf{S}^a- \mu ) + T\sum_{a=1}^A\mathsf{S}^a\mathcal{J} ^a,
\]
in accordance with \eqref{S_dot_simple} and \eqref{I_simple_example}. From the entropy equation and the last relation in \eqref{resulting_conditions_simple}, we have the rate of internal entropy production of the system as
\begin{equation}\label{rate_int_ent_pro_simple}
I:=\dot \Sigma=\dot S -\sum_{a=1}^A\mathsf{S}^a\mathcal{J} ^a= \frac{1}{T}\sum_{a=1}^A \mathcal{J}^a\left(\mathsf{H}^a-T\mathsf{S}^a- \mu \right).
\end{equation}

To summarize, from the variational formulations \eqref{VCond_simple}--\eqref{VCons_simple}, we have obtained the following evolution equations for the open system with $A$ ports:
\begin{equation}\label{evo_eqn_nh1}
\left\{
\begin{array}{l}
\vspace{0.2cm}\displaystyle \dot N= \sum_{a=1}^A\mathcal{J} ^a,\\
\displaystyle T\dot S= \sum_{a=1}^A \mathcal{J} ^a( \mathsf{H} ^a- T\mathsf{S}^a- \mu ) + T\sum_{a=1}^A\mathsf{S}^a\mathcal{J} ^a.
\end{array}\right.
\end{equation}
From this system, the first law is computed as
\[
\frac{d}{dt} U  = \sum_{a=1}^A\mathcal{J}^a(\mu^a+ \mathsf{S}^aT^a)=\sum_{a=1}^A\mathcal{J}^a\mathsf{H}^a.
\]
Equations \eqref{evo_eqn_nh1} yield a system of two ODEs for the two variables $S(t)$ and $N(t)$, for given molar flow rates $\mathcal{J}^a(t)$. The evolution of the other variables, such as $p(t)$, $T(t)$, $\mu(t)$, can be computed from $S(t)$ and $N(t)$ by using the state equation. In some situations, the molar flow rates are chosen in order the system to respect some properties, such as, for example a constant pressure $p(t)=p_0$.

\begin{example}[Filling an Insulated Tank with Pressurized Gas]{\rm 
In order to illustrate our formalism, we consider an insulated tank containing an amount of $N$ moles of ideal gas at temperature $T$ and pressure $p$. We~assume that gas with temperature $T^1(t)$ and pressure $p^1(t)$, which~are given prescribed functions of time $t$, is entering the tank at a single port with the molar flow rate $\mathcal{J}^1(t)$,~(\cite{Fu2010}, Section 8).

From the fundamental state equation of the ideal gas $U=U(S,V,N)$, all the thermodynamic relations can be obtained. For instance, we have
\begin{equation*}
\begin{split}
&p\mathsf{V}=RT, \quad \mathsf{U}= c_V(T-T_{\rm ref})+\mathsf{U}_{\rm ref}, \quad \mathsf{S}=- R\ln \frac{p}{p_{\rm ref}}+c_p \ln \frac{T}{T_{\rm ref}}+\mathsf{S}_{\rm ref},\\
&\mathsf{H}=c_{p}(T-T_{\rm ref})+\mathsf{H}_{\rm ref}, \quad 
\mu=c_{p}(T-T_{\rm ref})+  R\ln \frac{p}{p_{\rm ref}}-Tc_p \ln \frac{T}{T_{\rm ref}}-(T-T_{\rm ref})\mathsf{S}_{\rm ref}+\mu_{\rm ref},
\end{split}
\end{equation*}
where $R$ is the universal gas constant, $c_V$, $c_p$ are the heat capacity coefficients at constant volume and constant pressure of the ideal gas, and the index $()_{\rm ref}$ indicates reference values. The constants $R$, $c_V$, and $c_p$ are expressed in units $\frac{\textrm{J}}{\textrm{K}\cdot\textrm{mole}}$. Note that $\mathsf{U}$, $\mathsf{V}$, $\mathsf{S}$ {denote} the molar quantities associated to {$U$, $V$, $S$}.

The above variational formulation yields the equations of motion
\[
\dot N= \mathcal{J}^1, \quad \dot S= I + \mathsf{S}^1\mathcal{J}^1,
\]
where, for an ideal gas, $I$ is computed as
\[
I= \frac{1}{T}\mathcal{J}^1\left(\mathsf{H}^1-T\mathsf{S}^1- \mu \right)=\mathcal{J}^1 \left[ \frac{c_p}{T}(T^1-T)-R \ln\frac{p}{p^1}+c_p \ln \frac{T}{T^1}\right].
\]
This expression recovers the entropy production for this system, see, e.g.,~(\cite{Fu2010} Equation (8.74)). It~is associated to the irreversible process of mixing ideal gas at different temperature and pressure.

By rewriting the entropy production as

\begin{equation}\label{I_example1}
I= \mathcal{J}^1\left[c_{p}T  \left( \frac{T^1}{T} -1 - \ln \frac{T^1}{T} \right) -  RT \ln\frac{p}{p^1}\right],
\end{equation}
we see that $I\geq 0$ for all $T$, $T^1$, and for $p_1\geq p$. The second law thus consistently requires that $p^1\geq p$.}
\end{example}

\subsection{Variational Formulation of Open Simple Systems with Mechanical Interactions}\label{subsec_mechanics}

We shall now extend the previous variational formulation to the case of an open system including exchange of mechanical momenta. 
Such a system is described by two kind of mechanical variables. Mechanical variables $q=(q^{1},...,q^{n})$ describe the motion of the mechanical devices involved in the system, while $x=(x^{1},...,x^{m})$ are mechanical variables which indicate the motion of the species of the system. Through each of the $A$ ports, we have the flow rates $\mathcal{J}^a$, $\mathcal{J}^a_x$, and $\mathcal{J}^a_S$, $a=1,...,A$, corresponding to exchange of matter, mechanical momentum, and entropy. We assume that the species flows through the $a$-th port at velocity $v^a=(v^{a,1},...,v^{a,m})$. We also assume that the mechanical components are subject to friction and external forces, respectively, denoted by $F_x^{\rm fr}$, $F_q^{\rm fr}$ and $F_x^{\rm ext}$, $F_q^{\rm ext}$.

\medskip

\noindent{\bf Variational formulation.} We consider the following variational formulation:
\begin{equation}\label{VCond_simple_mech}
\begin{split}
&\delta \int_{t _1 }^{ t _2} \Big[ L(q, \dot q, x, \dot x, S, N)+\dot W N +(S- \Sigma ) \dot \Gamma \Big] dt \\
&\hspace{1.5cm}+\int_{t _1 }^{ t _2} \left[\big\langle F^{\rm ext}_q, \delta q \big\rangle+\big\langle F^{\rm ext}_x, \delta x \big\rangle  \right]dt=0 , \quad\;\;\; \textsc{Variational Condition}
\end{split}
\end{equation}
where the curves satisfy the 
nonlinear nonholonomic constraint 
\begin{equation}\label{PC_simple_mech}
\begin{split}
&\frac{\partial L}{\partial S}\dot \Sigma   = \big\langle F^{\rm fr}_q, \dot  q \big\rangle+\big\langle F^{\rm fr}_x, \dot  x \big\rangle + \sum_{a=1}^A \Big[ \big\langle\mathcal{J}_x ^a,  \dot x-v ^a \big\rangle+\mathcal{J} ^a\big( \dot W- \mu ^a\big) + \mathcal{J} ^a _S \big( \dot \Gamma - T ^a\big)\Big], \\
&\hspace{4cm}\textsc{{Kinematic} Constraint}
\end{split}
\end{equation}
and with respect to variations subject to the constraint
\begin{equation}\label{VCons_simple_mech}
\begin{split}
&\frac{\partial L}{\partial S}\delta \Sigma   = \big\langle F^{\rm fr}_q, \delta  q \big\rangle+\big\langle F^{\rm fr}_x, \delta  x \big\rangle + \sum_{a=1}^A \Big[ \big\langle\mathcal{J}_x^a, \delta x\big\rangle+ \mathcal{J} ^a\delta W +\mathcal{J} ^a _S\delta\Gamma  \Big]\\
&\hspace{2.5cm}\textsc{Variational Constraint}
\end{split}
\end{equation}
with $ \delta x(t_1)=\delta x(t_2)=0$, $\delta q(t_1)=\delta q(t_2)=0$ , $ \delta W(t_1)=\delta W(t_2)=0$. 

\medskip

\noindent{\bf Application of the variational formulation.}
By taking the variation of the integral in \eqref{VCond_simple_mech} and using the variational constraint \eqref{VCons_simple_mech}, we get the following conditions:
\begin{equation}\label{resulting_conditions_mech}
\begin{aligned}
\delta q:&\quad\frac{d}{dt}\frac{\partial L}{\partial \dot q^{i}}- \frac{\partial L}{\partial  q^{i}} = (F^{\rm fr}_q)_{i}+(F^{\rm ext}_q)_{i},\quad i=1,...,n, \\
\delta x:&\quad\frac{d}{dt}\frac{\partial L}{\partial \dot x^{r}}- \frac{\partial L}{\partial  x^{r}} = (F^{\rm fr}_x)_{r}+\sum_{a=1}^A(\mathcal{J}^a_x)_{r}+(F^{\rm ext}_x)_{r}, \quad r=1,...,m,\\
\delta S:&\quad\dot \Gamma = -\frac{\partial L}{\partial S}, \\
\delta W:&\quad\dot N= \sum_{a=1}^A\mathcal{J} ^a, \\
\delta N:&\quad\dot W=  -\frac{\partial L}{\partial N}, \\
\delta \Gamma :&\quad\dot S=\dot \Sigma +\sum_{a=1}^A\mathcal{J}^a_S.
\end{aligned}
\end{equation}
Extending the earlier definitions, we define the temperature and the chemical potential of the system in terms of the Lagrangian as $T:=-\frac{\partial L}{\partial S}$ and  $\mu:=-\frac{\partial L}{\partial N}$.
By inserting the conditions \eqref{resulting_conditions_mech} into the phenomenological constraint \eqref{PC_simple_mech}, we get the following entropy equation\vspace{-6pt}
\begin{equation}\label{S_dot_mech}
T\dot S=- \big\langle F^{\rm fr}_q, \dot q \big\rangle - \big\langle F^{\rm fr}_x, \dot x \big\rangle +\sum_{a=1}^A\Big[\big<\mathcal{J}^a_x, v^a-v \big> +\mathcal{J}^a(\mu^a-\mu) +\mathcal{J}_S^a(T^a-T)\Big]+ T\sum_{a=1}^A\mathcal{J} ^a_S.
\end{equation}

To summarize, from the variational formulations \eqref{VCond_simple_mech}--\eqref{VCons_simple_mech}, we have obtained the following evolution equations for the open system with mechanical interaction and $A$ ports:
\[
\left\{
\begin{array}{l}
\vspace{0.2cm}\displaystyle \frac{d}{dt}\frac{\partial L}{\partial \dot q^{i}}- \frac{\partial L}{\partial  q^{i}} = (F^{\rm fr}_q)_{i}+(F^{\rm ext}_q)_{i},\quad i=1,...,n, \\
\vspace{0.2cm}\displaystyle\frac{d}{dt}\frac{\partial L}{\partial \dot x^{r}}- \frac{\partial L}{\partial  x^{r}} = (F^{\rm fr}_x)_{r}+\sum_{a=1}^A(\mathcal{J}^a_x)_{r}+(F^{\rm ext}_x)_{r}, \quad r=1,...,m,\\
\vspace{0.2cm}\displaystyle \dot N= \sum_{a=1}^A\mathcal{J} ^a,\\
\displaystyle T\dot S=- \big\langle F^{\rm fr}_q, \dot q \big\rangle - \big\langle F^{\rm fr}_x, \dot x \big\rangle +\sum_{a=1}^A\Big[\big<\mathcal{J}^a_x, v^a-v \big> +\mathcal{J}^a(\mu^a-\mu) +\mathcal{J}_S^a(T^a-T)\Big]+ T\sum_{a=1}^A\mathcal{J} ^a_S.
\end{array}\right.
\]
This system of ODEs completely determines the time evolution of all the variables. In absence of thermodynamic effects, it reduces to the Euler-Lagrange equations of classical mechanics. 

Defining the total energy of the system as
\[
E(x, \dot x, q, \dot q, S,N):= \left\langle \frac{\partial L}{\partial \dot q}, \dot q \right\rangle +\left\langle \frac{\partial L}{\partial \dot x}, \dot x\right\rangle -L,
\]
we have the energy balance
\begin{equation}\label{energy_balance_mech}
\frac{d}{dt}E= \underbrace{\big\langle F^{\rm ext}_q, \dot q \big\rangle+ \big\langle F^{\rm ext}_x, \dot x \big\rangle}_{=P^{\rm ext}_W} + \underbrace{\sum_{a=1}^A\Big[\big\langle \mathcal{J}^a_x , v^a\big\rangle+ \mathcal{J}^a \mu^a+\mathcal{J}^a_S T^a \Big]}_{=P_M^{\rm ext}}.
\end{equation}

\begin{example}[{Flowing Fluid Carrying Kinetic Energy into a Piston Device}]{\rm We consider a piston with mass $M$ moving in a cylinder containing a species with internal energy $U=U(S,V,N)$. We assume that the cylinder has $A$ ports through which species is injected into or flows out of the cylinder with molar flow rates $\mathcal{J}^a$ and velocities $v^a$. The momentum and entropy flow rates accompanying these exchanges of species are given by

\[
\mathcal{J}^a_x= M_0\mathcal{J}^av^a \quad\text{and}\quad \mathcal{J}_S^a=\mathsf{S}^a\mathcal{J}^a,
\]
where $M_0$ is the molecular weight of the {species}.

We assume that the species flows through the $a$-th port and in the cylinder in a one dimensional direction, i.e., $v^a$ is parallel to $v=\dot x$, for each $a=1,...,A$, as illustrated in Figure~\ref{flowing_piston}.
The variable $q$ characterizes the one-dimensional motion of the piston, such that the volume occupied by the species is $V=\mathcal{A}q$, with $\mathcal{A}$ the sectional area of the cylinder. The Lagrangian of the system is the sum of the kinetic energy of the piston and the species, to which is subtracted the internal energy of the species
\[
L(q, \dot q, x, \dot x, S, N)= \frac{1}{2}M\dot q ^2+\frac{1}{2}M_0N \dot x ^2 -U(S,\mathcal{A}q,N).
\]
By a direct application of the variational formulation, the evolution equations are obtained as
\begin{equation}\label{equation_motion_2}
\dot N=\sum_{a=1}^A\mathcal{J}^a, \qquad M\ddot q - p\mathcal{A}= F^{\rm fr}_q+F^{\rm ext}_q, \qquad \frac{d}{dt} M_0N\dot x =F^{\rm fr}_x+ \sum_{a=1}^A\mathcal{J}^a_x+F^{\rm ext}_x, 
\end{equation}
where $p=-\frac{\partial U}{\partial V}$ is the pressure, together with the entropy Equation \eqref{S_dot_mech} which yields 
\begin{align*}
T\dot S&= - F^{\rm fr}_q\dot q- F^{\rm fr}_x \dot x  +\sum_{a=1}^A \left[ M_0v^a(v^a-v)+(\mu^a-\mu)+\mathsf{S}^a(T^a-T)\right]\mathcal{J}^a+T\sum_{a=1}^A\mathsf{S}^a\mathcal{J}^a\\
&= - F^{\rm fr}_q \dot q- F^{\rm fr}_x \dot x  +\sum_{a=1}^A \left[ \frac{1}{2}M_0(v^a)^2+ \frac{1}{2} M_0v^2- M_0 v^av + \mu_U^a+\mathsf{S}^aT^a- \mu_U-\mathsf{S}^a T\right]\mathcal{J}^a+T\sum_{a=1}^A\mathsf{S}^a\mathcal{J}^a\\
&= - F^{\rm fr}_q \dot q- F^{\rm fr}_x \dot x  +\sum_{a=1}^A \left[ \frac{1}{2}M_0(v^a-v)^2+\mathsf{H}^a-T\mathsf{S}^a- \mu_U \right]\mathcal{J}^a+T\sum_{a=1}^A\mathsf{S}^a\mathcal{J}^a,
\end{align*}
where we used the expression of the temperature $T:=-\frac{\partial L}{\partial S}=\frac{\partial U}{\partial S}$ and the expression of the chemical~potential

\[
\mu:=-\frac{\partial L}{\partial N}= - \frac{1}{2}M_0\dot x ^2+ \mu_U, \qquad \mu_U:=\frac{\partial U}{\partial N}.
\]
By translational invariance, the frictions forces verify $F^{\rm fr}_q=- F^{\rm fr}_x$, thus from this equation and the last equation in \eqref{resulting_conditions_mech}, we get the rate of entropy production within the system as

\begin{equation}\label{piston_I}
I:=\dot \Sigma= - \frac{1}{T}F^{\rm fr}_q  (\dot q-  \dot x)  +\frac{1}{T}\sum_{a=1}^A  \mathcal{J}^a\frac{1}{2}M_0(v^a-v)^2+\frac{1}{T}\sum_{a=1}^A\mathcal{J}^a\left[\mathsf{H}^a-T\mathsf{S}^a- \mu_U \right],
\end{equation}
where the first term represents the entropy production associated to the friction experiencing by the moving piston and the moving species, the second term is the entropy production associated to the mixing of gas at different velocities, and the last term is the entropy production associated to the mixing of gas at different pressure, and temperature. 
The second law requires that each of these terms is positive. Concerning the first term, the friction force must be dissipative and can be chosen as $F_q^{\rm fr}=- \lambda (\dot q-\dot x)$, where the phenomenological coefficient $\lambda(q,x,S)$ is determined experimentally, see~\cite{Gr1999}.
The second term is positive as long as $\mathcal{J}^a>0$ and it is zero if $\mathcal{J}^a<0$ since in this case $v^a=v$. Finally, the last term has been already analyzed in the previous example in the case of the ideal gas, see \eqref{I_example1}.

The energy balance is deduced from \eqref{energy_balance_mech} as
\begin{equation}\label{piston_energy}
\frac{d}{dt}E=   F^{\rm ext}_q \dot q  + F^{\rm ext}_x \dot x   + \sum_{a=1}^A\mathcal{J}^a\left[\frac{1}{2}M_0(v^a) ^2+\mathsf{H}^a\right]= P^{\rm ext}_W  + \underbrace{\sum_{a=1}^A\mathcal{J}^a\left[\mathsf{E}^a+p^a\mathsf{V}^a\right]}_{=P^{\rm ext}_M},
\end{equation}
where $\mathsf{E}^a= \frac{1}{2}M_0(v^a)^2 +\mathsf{U}^a$ denotes the total (kinetic plus internal) molar energy of the species flowing at the $a$-th port. This balance of energy is consistent with the general expression of the first law in~\eqref{law1_explicit}. Potential energies can be easily included in this example. External heating will be considered in~Section~\ref{sec_heat_matter}.
\begin{figure}[h!]
\begin{center}
\includegraphics[scale=0.9]{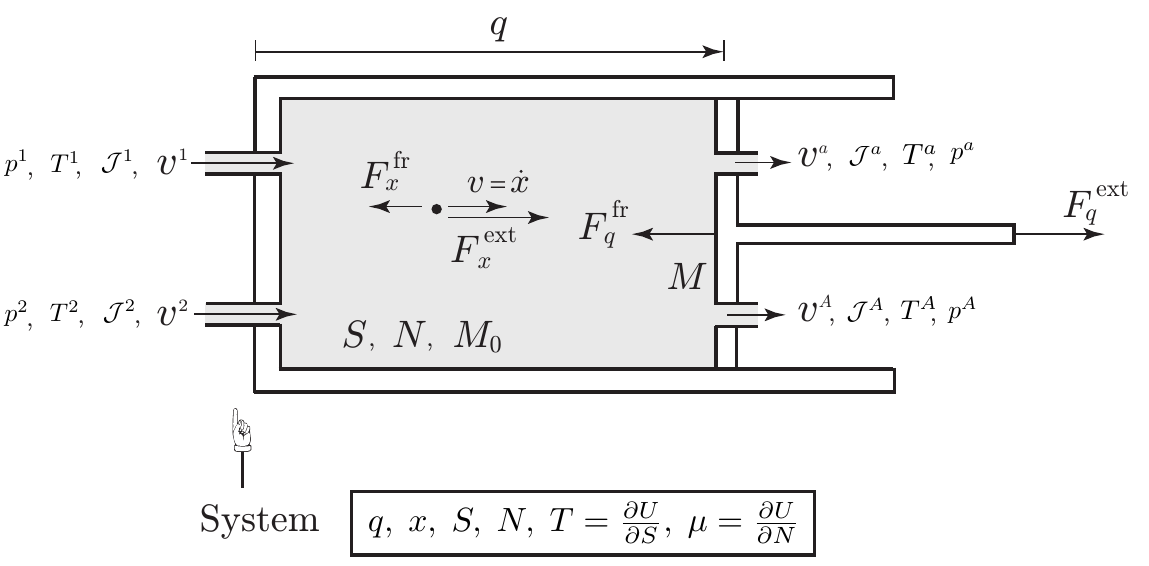}
\caption{Simple open system given by a flowing fluid carrying kinetic energy into a piston device.}
\label{flowing_piston}
\end{center}
\end{figure}}
\end{example}

\subsection{Variational Formulation of an Open Simple System with  Internal  Matter  Diffusion}\label{subsec_diffusion}

In this subsection, we extend the previous variational formulation to the case of open discrete systems experiencing internal diffusion processes. Diffusion is particularly important in biology where many processes depend on the transport of chemical species through bodies.

\begin{figure}[h!]
\begin{center}
\includegraphics[scale=0.6]{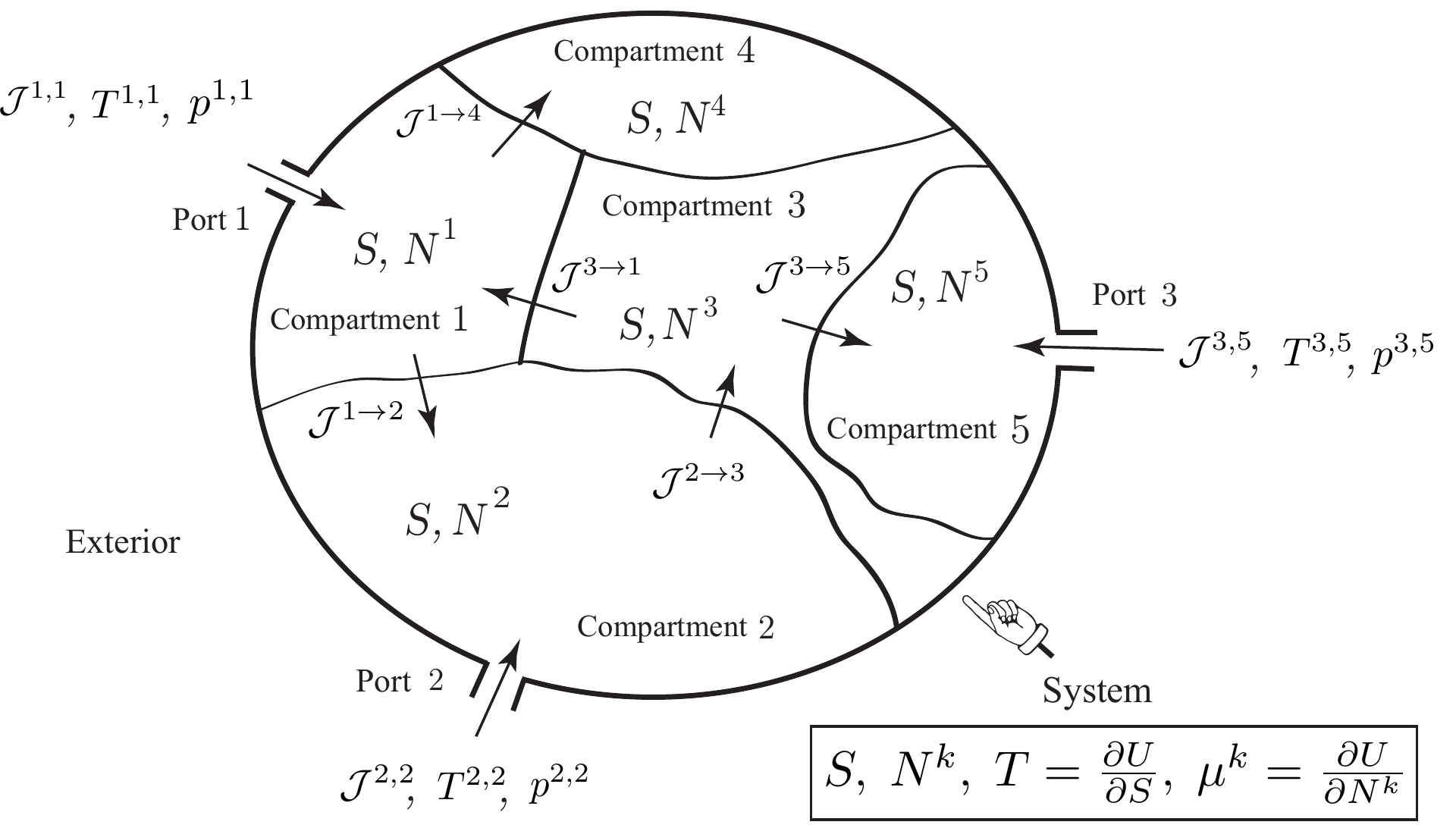}
\caption{Simple open system with a single species experiencing internal diffusion between several compartments, and transfer into or out of the system through several ports.}
\label{OpenSimpleDiscreteSystem}
\end{center}
\end{figure}

As illustrated in Figure~\ref{OpenSimpleDiscreteSystem}, we consider a thermodynamic system made of $K$ compartments that can exchange matter by diffusion across their common boundaries. 
Some compartments may have external ports, through which species can flow into or out of the compartment. We denote by $A_k$ the number of ports of compartment $k$, and by $\mathcal{J}^{a,k}$, the molar flow rate flowing into or out of compartment $k$ through the $a$-th port. We assume that the system has a single species and denote by $N ^k $ the number of moles of the species in the $k$-th compartment, $k=1,...,K$. We assume that the thermodynamic system is simple; i.e., a uniform entropy $S$, the entropy of the system, can be attributed to all the compartments. 
As~earlier, the thermodynamic quantities known at the ports are usually the pressure and the temperature $p^{a,k}$, $T^{a,k}$, from which the other thermodynamic quantities, such as $\mu^{a,k}$ or $\mathsf{S}^{a,k}$, are~deduced from the state equations of the gas.

The setting that we develop is well appropriate for the description of diffusion across composite membranes, e.g., composed of different elements arranged in a series or parallel array, which occur frequently in living systems and have remarkable physical properties, see~\cite{KeKa1963a,KeKa1963b,KeKa1963c,OsPeKa1973}.

For each compartment $k=1,...,K$, the mole balance equation is
\[
\frac{d}{dt} N ^k = \sum_{l=1}^K \mathcal{J} ^{l \rightarrow k}+  \sum_{a=1}^{A_k}\mathcal{J} ^{a,k},
\]
where $\mathcal{J} ^{ l \rightarrow k}=- \mathcal{J} ^{k \rightarrow l}$ is the molar flow rate from compartment $l$ to compartment $k$ due to diffusion of the species, and $\mathcal{J} ^{a,k}$ is the molar flow rate flowing into or out of compartment $k$ through the $a$-th port. 

\medskip

\noindent{\bf Lagrangian variational formulation.} 
By extending the variational formulation developed in Section~\ref{sec_foundations}, we~consider an open simple system with Lagrangian
\[
L(q, \dot q, S, N^1,...,N^K ),
\]
where $q=(q^{1},...,q^{n})$ are the mechanical variables of the system, $S$ the single entropy, and $N ^k$ the number of moles in the $k$-th compartment, $k=1,...,K$. The additional mechanical interactions described in Section~\ref{subsec_mechanics}, via the motion $x$ of the species, can be easily included in the present setting, too.

Given the external and friction forces $F^{\rm ext}$, $F^{\rm fr}$, the molar flow rates $\mathcal{J}^{l \rightarrow k}$, $\mathcal{J}^{a, k}$, and the entropy flow rates $\mathcal{J}^{l \rightarrow k}_S$, $\mathcal{J}^{a, k}_S$, the variational formulation is expressed as
\begin{equation}\label{VCond_simple_diffusion}
\begin{split}
&\delta \int_{t _1 }^{ t _2} \Big[ L(q, \dot q, S, N^1,...,N^K)+\sum_{k=1}^{K}\dot W_kN^k +(S-\Sigma)\dot\Gamma\Big] dt + \int_{t_1}^{t_2} \big<F^{\rm ext }, \delta q \big>dt=0 ,\\
&\hspace{4cm}\textsc{Variational Condition}
\end{split}
\end{equation}
where the curves satisfy the 
nonlinear nonholonomic constraint
\begin{equation}\label{PC_simple_diffusion}
\begin{split}
&\frac{\partial L}{\partial S}\dot\Sigma  = \big\langle F^{\rm fr}, \dot q \big\rangle + \sum_{k,l=1}^{K}  \mathcal{J}^{l \rightarrow k} \dot W _k+\sum_{k=1}^K\sum_{a=1}^{A_k}\Big[\mathcal{J}^{a, k}(\dot W^ k- \mu^{a,k})+\mathcal{J}^{a, k}_S(\dot \Gamma- T^{a,k})\Big],\\
&\hspace{3cm}\textsc{{Kinematic} Constraint}
\end{split}
\end{equation}
and with respect to variations subject to the constraint
\begin{equation}\label{VC_simple_diffusion}
\begin{split}
&\frac{\partial L}{\partial S}\delta\Sigma  = \big\langle F^{\rm fr}, \delta q \big\rangle + \sum_{k,l=1}^{K}  \mathcal{J}^{l \rightarrow k} \delta W _k+\sum_{k=1}^K\sum_{a=1}^{A_k} \Big[\mathcal{J}^{a, k}\delta W^ k+\mathcal{J}^{a, k}_S\delta \Gamma \Big],\\
&\hspace{3cm}\textsc{Variational Constraint}
\end{split}
\end{equation}
with $\delta q(t_1)=\delta q(t_2)=0$ and $ \delta W_k(t_1)=\delta W_k(t_2)=0$.

By taking the variation of the integral in \eqref{VCond_simple_diffusion} and using the variational constraint \eqref{VC_simple_diffusion}, we get the following conditions:
\begin{equation}
\begin{aligned}\label{OpenSystemEquation} 
\delta q^{i}:&\quad \frac{d}{dt}\frac{\partial L}{\partial \dot{q}^{i}}-\frac{\partial L}{\partial q^{i}}=(F^{\rm fr})_{i}+(F^{\rm ext })_{i}, \quad i=1,...,n,\\
\delta S:&\quad\dot \Gamma = -\frac{\partial L}{\partial S}, \\
\delta N^k:&\quad  \dot W_{k}=-\frac{\partial L}{\partial N^{k}},\quad k=1,...,K\\
\delta W_k:&\quad  \frac{d}{dt} N ^k = \sum_{l=1}^K \mathcal{J} ^{l \rightarrow k}+  \sum_{a=1}^{A_k}\mathcal{J} ^{a,k}, \quad k=1,...,K,\\
\delta \Gamma :&\quad\dot S=\dot \Sigma +\sum_{k=1}^K\sum_{a=1}^{A_k}\mathcal{J}^{a,k}_S.
\end{aligned}
\end{equation}
As before, we define the temperature and the chemical potential potentials of each compartment as $T:=-\frac{\partial L}{\partial S}$ and $\mu^k:=-\frac{\partial L}{\partial N^k}$. By inserting the conditions \eqref{OpenSystemEquation} into the phenomenological constraint \eqref{PC_simple_diffusion} and using 
\[
\sum_{k,l=1}^{K}\mathcal{J} ^{l \rightarrow k}\mu^k= \sum_{k<l}
\mathcal{J} ^{l \rightarrow k} (\mu^k -\mu^l),
\]
we get the following entropy equation
\[
T\dot S=-\big< F^{\rm fr}, \dot{q}\big> +\sum_{k<l}
\mathcal{J} ^{k \rightarrow l} (\mu^k -\mu^l)+\sum_{k=1}^K\sum_{a=1}^{A_k} \left[\mathcal{J}^{a, k}(\mu^{a,k}-\mu^ k)+\mathcal{J}^{a, k}_S(T^{a,k}-T)\right]+T\sum_{k=1}^K\sum_{a=1}^{A_k}\mathcal{J}^{a,k}_S.
\]
The rate of entropy production within the system is given by
\[
I:=\dot\Sigma=  - \frac{1}{T}\big< F^{\rm fr}, \dot{q}\big> +\frac{1}{T} \sum_{k<l}
\mathcal{J} ^{k \rightarrow l} (\mu^k -\mu^l)+\frac{1}{T}\sum_{k=1}^K\sum_{a=1}^{A_k} \Big[\mathcal{J}^{a, k}(\mu^{a,k}-\mu^ k)+\mathcal{J}^{a, k}_S(T^{a,k}-T)\Big],
\]
where the three terms correspond to the entropy production associated to mechanical friction, diffusion, and mixing of gas flowing into the system through the ports. 

By the second law, the entropy production must be always positive and hence suggests the phenomenological relations
\[
F^{\rm fr}_{i}=-\lambda_{ij} \dot{q}^{j}\;\; \text{and}\;\; \mathcal{J} ^{k \rightarrow l}=G^{kl} (\mu ^k-\mu^l),
\]
where $\lambda_{ij},\;i,j=1,...,n$ and $G^{kl},\; k,l=1,...,K$ are functions of the state variables, with $\lambda_{ij}$ positive semi-definite and $G^{kl}\geq 0$, for all $k,l$. The third term has been already discussed earlier.

The total energy associated to the Lagrangian $L(q, \dot q, S, N^1,...,N^K)$ is defined as
\[
E(q, \dot q, S,N^1,...,N^K):= \left\langle \frac{\partial L}{\partial \dot q}, \dot q \right\rangle -L,
\]
and satisfies the energy balance
\[
\frac{d}{dt}E=\left< F^{\rm ext}, \dot{q} \right>+\sum_{k=1}^{K} \sum_{a=1}^{A_k}\Big[
\mathcal{J} ^{a,k}\mu^{a,k}+ \mathcal{J} ^{a,k}_ST^{a,k}\Big]=P^{\rm ext}_W+P^{\rm ext}_M.
\]

\subsection{Relation with the Abstract Variational Formulation}\label{link_abstract}

We shall here quickly mention how the variational formulations \eqref{VCond_simple}--\eqref{VCons_simple}, as well as its more general versions \eqref{VCond_simple_mech}--\eqref{VCons_simple_mech} and
\eqref{VCond_simple_diffusion}--\eqref{VC_simple_diffusion}, fit into the abstract variational formulation developed in~\eqref{Vcond_abstract}--\eqref{PC_abstract}.

Recall that the abstract variational formulation consists of two kind of constraints, the variational constraint $\widehat{C}_{V}$ in \eqref{VC_abstract} and the kinematic constraints $C_{K}$ in \eqref{PC_abstract}, which are related with each other in a very specific way.
Concerning the variational formulation \eqref{VCond_simple_mech}--\eqref{VCons_simple_mech}, for instance, the curve $q(t)$ of the abstract setting is given by the collection of curves $(q(t),x(t),S(t),N(t),\Gamma(t),W(t),\Sigma(t))$, and the Lagrangian $L(t,q,\dot q)$ of the abstract setting is time independent and given by the function $L(q, \dot q, x, \dot x, S, N)+\dot W N +(S- \Sigma ) \dot \Gamma$. The linear function $ \delta q^i\mapsto A_i(t,q,\dot q)^r \delta q_i$ is given by the linear map
\[
(\delta q, \delta x, \delta S, \delta N, \delta \Gamma, \delta W, \delta\Sigma)\mapsto - \frac{\partial L}{\partial S}\delta \Sigma  + \big\langle F^{\rm fr}_q, \delta  q \big\rangle+\big\langle F^{\rm fr}_x, \delta  x \big\rangle + \sum_{a=1}^A \Big[ \big\langle\mathcal{J}_x^a, \delta x\big\rangle+ \mathcal{J} ^a\delta W +\mathcal{J} ^a _S\delta\Gamma  \Big],
\]
where one notes that the explicit time dependence may come from the time dependence of the functions $\mathcal{J}_x^a$, $\mathcal{J}^a$, $\mathcal{J}_p^a$. On the other hand, the functions $B^r(t,q, \dot q)$ in \eqref{PC_abstract} correspond to
\[
-\sum_{a=1}^A \Big[ \big\langle\mathcal{J}_x^a, v^a\big\rangle+ \mathcal{J} ^a\mu^a +\mathcal{J} ^a _ST^a \Big].
\]
Then, one notices that with these choices, the variational constraint $\widehat{C}_V$ in \eqref{VC_abstract} yields \eqref{VCons_simple_mech}, and the kinematic constraint $C_K$ in \eqref{PC_abstract} yields \eqref{PC_simple_mech}. Similarly, the variational formulations in \eqref{VCond_simple_diffusion}--\eqref{VC_simple_diffusion} can be also seen as a special instance of the abstract formulations  \eqref{VCond_simple}--\eqref{VCons_simple}.

\color{black}

\subsection{Examples of Open Simple Systems with Matter Transfer}\label{subsec_examples}

Here we consider three examples of open simple systems with matter diffusion of a single chemical species. The first example is a simple system with matter diffusion between two compartments. The~other two examples are diffusion processes appearing in composite membranes, consisting of serial or parallel interconnections of membranes~(\cite{KeKa1963a,KeKa1963b,KeKa1963c};~\cite{OsPeKa1973}, Section 4.7).

\begin{example}[Diffusion between Two Compartments]
{\rm Consider an open system with matter diffusion of a single species between two compartments and transfer of matter into or out of the compartments via two external ports with molar flow rate $\mathcal{J} ^{a_{k}}$, temperature $T^{a_{k}}$ and pressure $p^{a_k},\;k=1,2$ as shown in Figure~\ref{simple_mass_transfer}. In this case, the Lagrangian is just given by the internal energy of the system, \mbox{$L(q, \dot q, S, N^1,N^2 )=-U(S,N^1,N^2)$}.  From the above developments it follows the evolution equations as
\begin{equation}\label{SimpleMassTransferEqn} 
\left\{ 
\begin{array}{l}
\displaystyle \dot N^1=  \mathcal{J} ^{ 2 \rightarrow 1}+\mathcal{J} ^{a_1},\qquad \dot N^2=  \mathcal{J} ^{ 1 \rightarrow 2}+\mathcal{J} ^{a_2},\\[1mm]
\displaystyle T\dot S=\mathcal{J} ^{1 \rightarrow 2} (\mu _1-\mu_{2}) + \sum_{k=1}^2\mathcal{J}^{a_k}(\mathsf{H}^{a_k}-T\mathsf{S}^{a_k}-\mu^k)+T \sum_{k=1}^2\mathsf{S}^{a_k}\mathcal{J}^{a_k},
\end{array}
\right.
\end{equation} 
with $T=\frac{\partial U}{\partial S}$ and $\mu_{k}=\frac{\partial U}{\partial N^{k}},\;k=1,2$, together with the relations
\[
T=\dot \Gamma, \quad \mu^{k}=\dot W^{k},\quad k=1,2,
\]
where we introduced the molar enthalpy at the ports $a_k$, $k=1,2$, as
\[
\mathsf{H}^{a_k}=\mathsf{U}^{a_k}+p^{a_k}\mathsf{V}^{a_k}= \mu^{a_k}+T^{a_k}\mathsf{S}^{a_k},\quad k=1,2.
\]
In \eqref{SimpleMassTransferEqn}, the first two equations are the mole balance in each compartment, where $\mathcal{J}^{1 \rightarrow 2}$ is the molar flow rate due to diffusion between the compartments. The last equation is the entropy equation, where~we note that the rate of entropy production within the system reads
\[
I:=\dot\Sigma= \frac{1}{T}\left[\mathcal{J} ^{1 \rightarrow 2} (\mu _1-\mu_{2}) +  \sum_{k=1}^2\mathcal{J}^{a_k}(\mathsf{H}^{a_k}-T\mathsf{S}^{a_k}-\mu^k)\right].
\]
From the second law, the internal entropy production must be always positive. In particular, for the process of diffusion, this suggests the phenomenological relation
\[
\mathcal{J} ^{1 \rightarrow 2}=G^{12}(\mu ^1-\mu^2),
\]
where $G^{12}\geq 0$ is a function of the state variables.

One verifies that along a solution curve $(S(t), N^{1}(t), N^{2}(t))$ of \eqref{SimpleMassTransferEqn}, we have the balance of energy
\begin{equation}\label{EnergyBalance}
\frac{d}{dt}U=\frac{\partial U}{\partial S}\dot{S} +\sum_{k=1}^{2}\frac{\partial U}{\partial N^{k}}\dot{N}^{k}
= \sum_{k=1}^{2}\mathsf{H}^{a_k}\mathcal{J} ^{a_k}=P_{M}^{\rm ext}.
\end{equation}

\begin{figure}[h!]
\begin{center}
\includegraphics[scale=0.7]{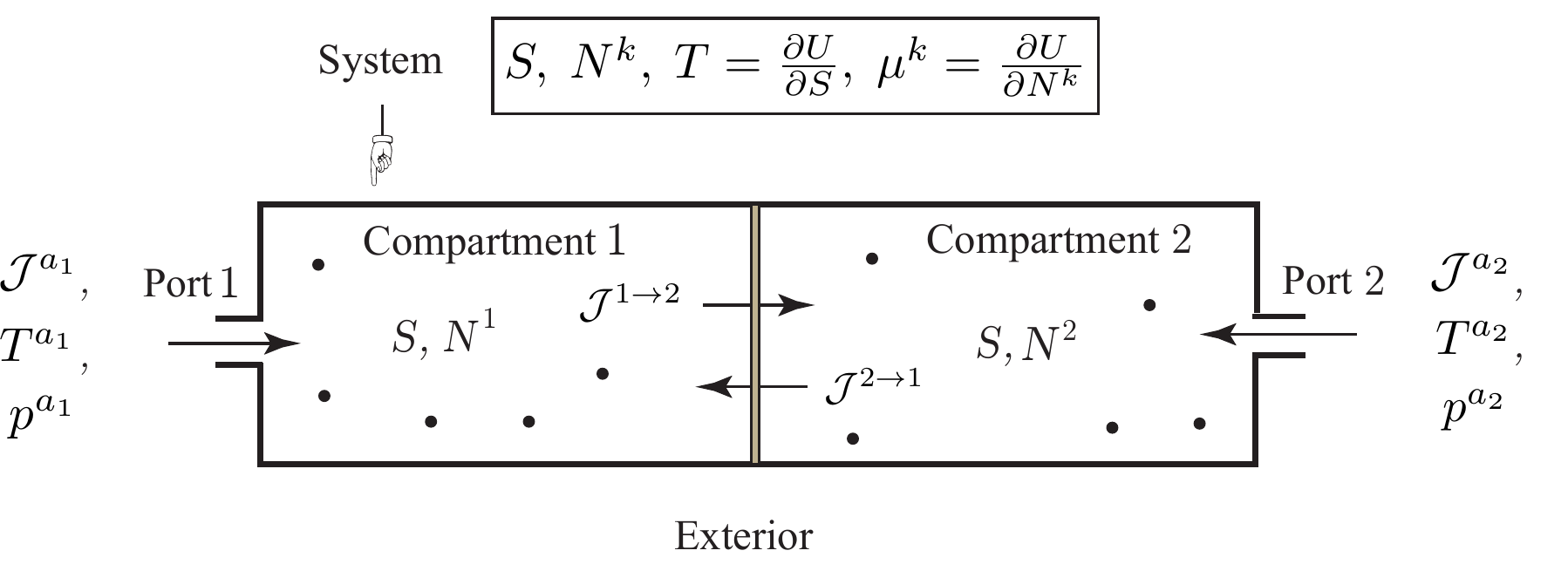}
\caption{Simple open system with a single chemical component experiencing diffusion between two compartments, and transfer into or out of the system through two ports.}
\label{simple_mass_transfer}
\end{center}
\end{figure}}
\end{example}

\begin{figure}[h!]
\begin{center}
\includegraphics[scale=0.6]{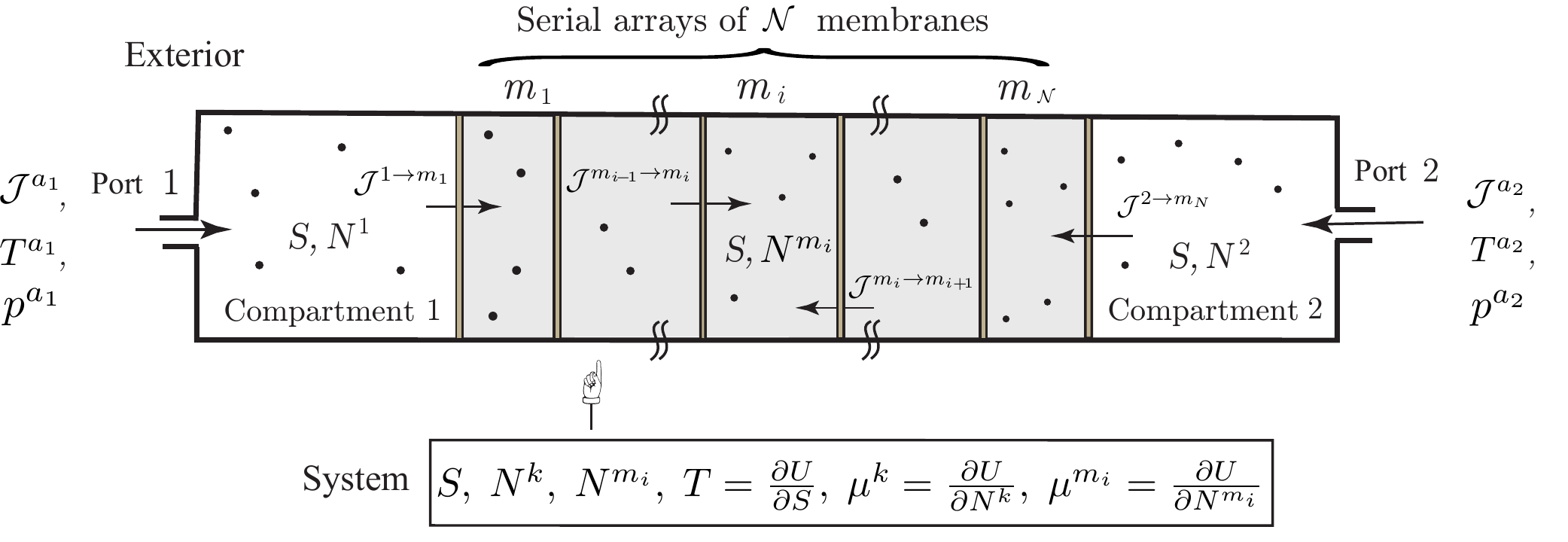}
\caption{Simple open system  with diffusion of a single chemical species through a composite membrane made of a serial array of $\mathcal{N}$ elements.}
\label{ser_membrane_mass_transfer}
\end{center}
\end{figure}

\begin{example}[{Diffusion through a Series Array of Membranes}]{\rm To illustrate our variational formulation, we consider the open system represented
on Figure~\ref{ser_membrane_mass_transfer}, in which diffusion occurs through a composite membrane made of different elements arranged in a series-array. In this example, matter also transfers into or out of the compartments via two external ports with molar flow rate $\mathcal{J} ^{a_{k}}$, temperature $T^{a_{k}}$ and pressure $p^{a_k},\;k=1,2$. In this case, the variational formulation yields the evolution equations
\[
\left\{ 
\begin{array}{l}
\displaystyle \dot N^1=  \mathcal{J} ^{ m_{1} \rightarrow 1}+\mathcal{J} ^{ a_1},\qquad \displaystyle \dot N^2=  \mathcal{J}^{m_{K} \rightarrow 2}+\mathcal{J} ^{ a_2},\\[3mm]
\displaystyle \dot N^{m_{i}}=  \mathcal{J}^{ m_{i-1} \rightarrow m_{i}}+ \mathcal{J}^{ m_{i+1} \rightarrow m_{i}},\quad i=1,...,\mathcal{N},\\[2mm]
\displaystyle T\dot S=\mathcal{J} ^{1 \rightarrow m_{1}} (\mu_{1}-\mu _{m_{1}})+\sum_{i=1}^{\mathcal{N}} \mathcal{J} ^{m_{i} \rightarrow m_{i+1}} (\mu_{m_{i}}-\mu _{m_{i+1}})+\mathcal{J} ^{2 \rightarrow m_{K}} (\mu_{2}-\mu _{m_{K}})\\[1mm]
\displaystyle\qquad\qquad+ \sum_{k=1}^2\mathcal{J}^{a_k}(\mathsf{H}^{a_k}-T\mathsf{S}^{a_k}-\mu^k)+T \sum_{k=1}^2\mathsf{S}^{a_k}\mathcal{J}^{a_k},
\end{array}
\right.
\]
together with the relations
\[
T=\dot \Gamma, \quad \mu^{k}=\dot W^{k},\quad \mu^{m_i}=\dot W^{m_i},
\]
where $\mathsf{H}^{a_k}$ is the molar enthalpy at the port $a_k$. From the second law, the entropy production must be always positive and hence dictates the choice of the phenomenological equations
\[
\mathcal{J} ^{1 \rightarrow m_{1}}=G^{1,m_{1}}(\mu_{1}-\mu _{m_{1}}),\;\;\;
\mathcal{J} ^{m_{i} \rightarrow m_{i+1}}=G^{m_{i},m_{i+1}} (\mu_{m_{i}}-\mu _{m_{i+1}}),\;\;\;
\mathcal{J} ^{2 \rightarrow m_{K}}=G^{2,m_{K}}(\mu_{2}-\mu _{m_{K}}),
\]
for positive state functions $G^{1,m_{1}}$, $G^{m_{i},m_{i+1}}$, $G^{2,m_{K}}$.
The energy balance equation holds as

\begin{equation}\label{EBE}
\frac{d}{dt}U= \sum_{k=1}^{2}\mathsf{H}^{a_k}\mathcal{J} ^{a_k}.
\end{equation}}
\end{example}

\begin{figure}[h!]
\begin{center}
\includegraphics[scale=0.56]{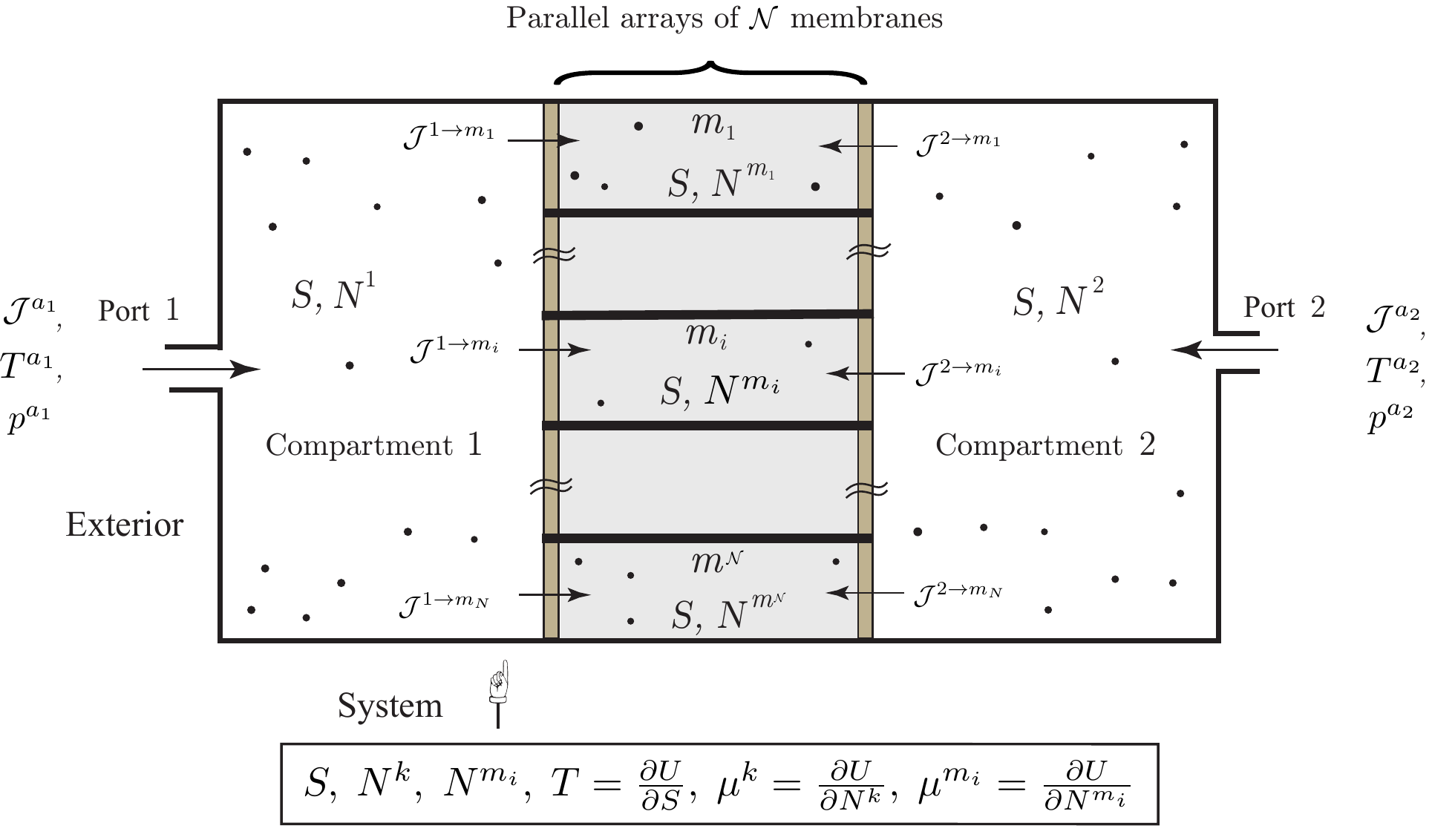}
\caption{Simple open system  with diffusion of a single chemical species through a composite membrane made of a parallel array of $\mathcal{N}$ elements.}
\label{par_membrane_mass_transfer}
\end{center}
\end{figure}

\begin{example}[Diffusion through a Parallel Array of Membranes]{\rm We now consider diffusion through a composite membrane made of different elements arranged in a parallel-array, as illustrated in Figure~\ref{par_membrane_mass_transfer}. As before, matter also transfers into or out of the compartments via two external ports. In this case, the variational formulation directly yields the evolution equations
\[
\left\{ 
\begin{array}{l}
\displaystyle \dot N^1=  \sum_{i=1}^{\mathcal{N}}\mathcal{J} ^{ m_{i} \rightarrow 1}+\mathcal{J} ^{ a_1},\qquad \dot N^2=  \sum_{i=1}^{\mathcal{N}} \mathcal{J}^{m_{i} \rightarrow 2}+\mathcal{J} ^{ a_2},\\[4mm]
\displaystyle \dot N^{m_{i}}=  \mathcal{J}^{1 \rightarrow m_{i}}+ \mathcal{J}^{2 \rightarrow m_{i}},\quad i=1,...,\mathcal{N},\\[2mm]
\displaystyle T\dot S=\sum_{i=1}^{\mathcal{N}} \left[\mathcal{J} ^{1 \rightarrow m_{i}} (\mu_{1}-\mu _{m_{i}})+\mathcal{J} ^{2 \rightarrow m_{i}} (\mu_{2}-\mu _{m_{i}})\right]+ \sum_{k=1}^2\mathcal{J}^{a_k}(\mathsf{H}^{a_k}-T\mathsf{S}^{a_k}-\mu^k)+T \sum_{k=1}^2\mathsf{S}^{a_k}\mathcal{J}^{a_k},
\end{array}
\right.
\]
together with the relations
\[
T=\dot \Gamma, \quad \mu^{k}=\dot W^{k}, \quad \mu^{m_i}=\dot W^{m_i}.
\]
As earlier, the second law suggests the phenomenological equations
\[
\mathcal{J} ^{1 \rightarrow m_{i}}=G^{1,m_{i}} (\mu_{1}-\mu _{m_{i}}),\quad
\mathcal{J} ^{2 \rightarrow m_{i}}=G^{2, m_{i}} (\mu_{2}-\mu _{m_{i}})
\]
for positive state functions $G^{1,m_{i}}$, $G^{2, m_{i}}$. The energy balance equation is the same as in \eqref{EBE}.}
\end{example}

\section{Open Discrete Systems with Heat and Matter Transfer}\label{sec_heat_matter}

In this section, we extend the variational formulation developed above to the case of {non-simple} open thermodynamic systems. Such systems are composed of several subsystems, each of them being simple and open, and exchanges heat and matter with other subsystems.

\subsection{Variational Formulation  for Discrete Open Systems}
\begin{figure}[h!]
\begin{center}
\includegraphics[scale=0.7]{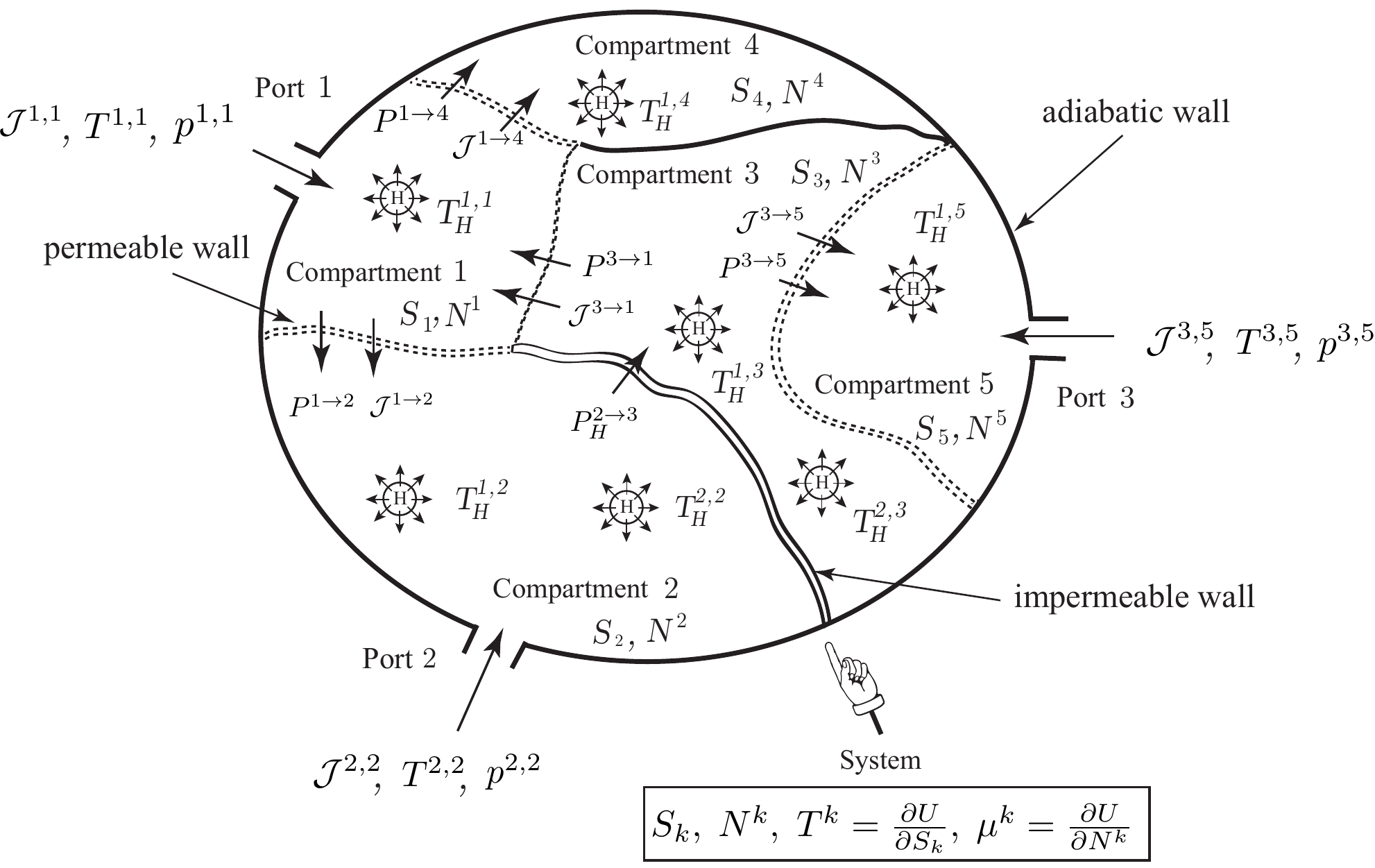}
\caption{Open system with a single species experiencing internal diffusion an heat transfer between several compartments, and transfer into or out of the system through several ports.}
\label{heat_mass_transfer}
\end{center}
\end{figure}

Consider an open system with a single species experiencing internal diffusion and heat transfer between several compartments, and transfer into or out of the system through several ports. The walls between compartments can be adiabatic, impermeable, or permeable. An example of such a system is illustrated in Figure~\ref{heat_mass_transfer}. {For simplicity, we do not consider the mechanical interactions. They can be included as in Section~\ref{subsec_mechanics}.}

We denote by $S^k$ and $N^k$  the entropies and the number of moles $N^k$ of the $k$-th compartment, $k=1,...,K$. We assume that the system also involves mechanical variables $q=(q^{1},...,q^{n})$.
We denote by $P^{k \rightarrow l}$ the total power exchange due to the heat and matter transfer between the $k$-th and $l$-th compartments and by $\mathcal{J}^{k \rightarrow l}$ the molar flow rate of the chemical species between the $k$-th and $l$-th compartments.
When the power exchange is only associated to heat transfer, it is denoted by $P_{H}^{k \rightarrow l}$.

We assume that the Lagrangian of the system is of the form
\[
L(q, \dot q, S_1,...,S_{K}, N^1,...,N^{K}),
\]
i.e., it depends on the mechanical variables $q$, as well as the number of moles and the entropies of each compartment. As earlier, we denote by $F^{\rm ext}$ and $F^{\rm fr}$ the external and friction forces. We also introduce the fluxes $J^{lk}$, $k\neq l$ such that $J^{kl}=J^{lk}$. The relation between the fluxes $J^{kl}$ and the total power exchanges $P^{k\rightarrow l}$ will be given later. For the purpose of the variational formulation, it is convenient to define the flux $J^{kl}$ for $k=l$ as
\[
J^{kk}:=- \sum_{l\neq k}J^{kl},
\]
so that we have 
\begin{equation}\label{propertyJ}
\sum_{k=1}^KJ^{kl}=0.
\end{equation}
As earlier, the $k$-th compartment may have $A_k$ ports, through which species can flow out or into the system.
In addition, we also assume that the $k$-th compartment may have $B_k$ heat sources of temperature $T^{b,k}_{H}$.

Given the Lagrangian $L$, the external and friction forces $F^{\rm ext}$, $F^{\rm fr}_{k}$, the fluxes $J^{kl}$, the molar flow rates $\mathcal{J}^{l \rightarrow k}$, $\mathcal{J}^{a, k}$, and the entropy flow rates $\mathcal{J}^{l \rightarrow k}_S$, $\mathcal{J}^{a, k}_S$, the variational formulation is expressed as
\begin{equation*}
\begin{split}
&\delta \int_{t _1 }^{ t _2} \Big[ L(q, \dot q, S_1,...,S_{K}, N^1,...,N^{K})+\sum_{k=1}^{K} \dot W_k N^k +\sum_{k=1}^{K} \dot{\Gamma }^k( S_k- \Sigma  _k)\Big] dt \\
&\hspace{5cm}+ \int_{t_1}^{t_2} \big<F^{\rm ext }, \delta q \big>dt =0,\qquad \textsc{Variational Condition}
\end{split}
\end{equation*}
where the curves satisfy the nonlinear nonholonomic constraint 
\begin{equation*}
\begin{split}
&\frac{\partial L}{\partial S_k}\dot { \Sigma }_k  = \big\langle F^{\rm fr}_k, \dot q \big\rangle + \sum_{l=1}^{K} J^{kl} \dot \Gamma ^{l}+ \sum_{l=1}^{K} \mathcal{J}^{l \rightarrow k} \dot W _k+\sum_{a=1}^{A_k}\Big[\mathcal{J}^{a, k}(\dot W^ k- \mu^{a,k})+\mathcal{J}^{a, k}_S(\dot \Gamma^k- T^{a,k})\Big]\\
&\hspace{4cm}+\sum_{b=1}^{B_k}\mathcal{J}^{b, k}_S(\dot \Gamma^k- T_{H}^{b,k}),\quad k=1,...,K, \qquad \textsc{{Kinematic} Constraint}
\end{split}
\end{equation*}
and with respect to variations subject to the constraint
\begin{equation*}
\begin{split}
&\frac{\partial L}{\partial S_k} \delta  { \Sigma }_k  = \big\langle F^{\rm fr}_k, \delta q \big\rangle + \sum_{l=1}^{K} J^{kl} \delta \Gamma ^{l}+ \sum_{l=1}^{K} \mathcal{J}^{l \rightarrow k} \delta  W _k+\sum_{a=1}^{A_k} \Big[\mathcal{J}^{a, k}\delta W^ k+\mathcal{J}^{a, k}_S\delta \Gamma^k \Big]\\
&\hspace{6cm}+\sum_{b=1}^{B_k}\mathcal{J}^{b, k}_S\delta \Gamma^k,\quad k=1,...,K,\qquad \textsc{Variational Constraint}
\end{split}
\end{equation*}
with $\delta q(t_1)=\delta q(t_2)=0$ and $ \delta W_k(t_1)=\delta W_k(t_2)=0$.

\medskip

Taking the variations, we get
\begin{align*} 
&\int_{t_1}^{t_2}\left[\left( \frac{\partial L}{\partial q}^{i}-\frac{d}{dt} \frac{\partial L}{\partial \dot q^{i}} \right) \delta q^{ i}+ \sum_{k=1}^{K} \frac{\partial L}{\partial S_k}\delta S_k + \sum_{k=1}^{K}\frac{\partial L}{\partial N^k} \delta N^k  - \sum_{k=1}^{K}\delta\Gamma ^k(\dot S_k- \dot \Sigma  _k)\right.\\
& \qquad\qquad\qquad\qquad  + \left.\sum_{k=1}^{K} \dot \Gamma ^k( \delta S_k- \delta \Sigma  _k)- \sum_{k=1}^{K} \delta W_k\dot N^k +\sum_{k=1}^{K} \dot W _k \delta N^k+  \big <F^{\rm ext }, \delta q \big>
\right] dt=0.
\end{align*} 
With the help of the variational constraint we replace $ \delta \Sigma _k$ and we get
\begin{align*} 
\delta q^{i}:&\quad\frac{\partial L}{\partial q^{i}}-\frac{d}{dt} \frac{\partial L}{\partial \dot q^{i}} -\sum_{k=1}^K \left(\frac{\dot \Gamma^k}{ \frac{\partial L}{\partial S_k} }F^{\rm fr}_k\right)_{i} +(F^{\rm ext })_{i}=0,\quad i=1,...,n,\\
\delta S_k:&\quad  \frac{\partial L}{\partial S_k}+ \dot\Gamma ^k=0, \quad k=1,...,K,\\
\delta N^k:&\quad\frac{\partial L}{\partial N^k}+ \dot W _k=0, \quad k=1,...,K,\\
\delta W_k:&\quad-\dot N^k -\frac{\dot \Gamma ^k}{ \frac{\partial L}{\partial S_k} }\sum_{l=1}^K \mathcal{J}^{l \rightarrow k} -\frac{\dot \Gamma ^k}{ \frac{\partial L}{\partial S_k} }\sum_{a=1}^{A_k} \mathcal{J}^{a,k}, \quad k=1,...,K,\\
\displaystyle\delta \Gamma ^k: &\quad- \dot S_k + \dot \Sigma  _k -\sum_{l=1}^K \frac{\dot \Gamma ^l}{ \frac{\partial L}{\partial S_l} }J^{lk}-\frac{\dot \Gamma ^k}{ \frac{\partial L}{\partial S_k} }\sum_{a=1}^{A_k}\mathcal{J}^{a,k}_S-\frac{\dot \Gamma ^k}{ \frac{\partial L}{\partial S_k} }\sum_{b=1}^{B_k}\mathcal{J}^{b,k}_S=0,\quad k=1,...,K.
\end{align*}
From these conditions, together with \eqref{propertyJ}, we get the evolution equations for the open system with internal diffusion and heat transfer as
\begin{equation}\label{EvolEqn_Open}
\left\{
\begin{array}{l}
\displaystyle\vspace{0.2cm}\frac{d}{dt} \frac{\partial L}{\partial \dot q^{i}}-  \frac{\partial L}{\partial q^{i}}= \sum_{k=1}^K(F^{\rm fr}_k)_{i} + (F^{\rm ext })_{i}, \quad i=1,...,n,\\
\displaystyle\vspace{0.2cm}\dot N^k= \sum_{l=1}^{K} \mathcal{J}^{l \rightarrow k}+\sum_{a=1}^{A_k} \mathcal{J}^{a,k}, \quad k=1,...,K,\\
\displaystyle \vspace{0.2cm} T^k \dot S_k= - \big\langle F^{\rm fr}_k, \dot q \big\rangle - \sum _{l=1}^K J^{kl}(T^{l}-T^{k}) -\sum_{l=1}^K \mathcal{J} ^{l \rightarrow k} \mu _k-\sum_{a=1}^{A_k}\Big[\mathcal{J}^{a, k}(\mu^ k- \mu^{a,k})+\mathcal{J}^{a, k}_S(T^k- T^{a,k})\Big]\\
\qquad\qquad \displaystyle-\sum_{b=1}^{B_k}\mathcal{J}^{b, k}_S(T^k- T_{H}^{b,k})+T_{k}\sum_{a=1}^{A_k}\mathcal{J}^{a,k}_S+T_{k}\sum_{b=1}^{B_k}\mathcal{J}^{b,k}_S, \quad k=1,...,K,
\end{array}
\right.
\end{equation}
together with the conditions
\[
\dot\Gamma ^k=-\frac{\partial L}{\partial S_k}=:T^k, \qquad \dot W_k=-\frac{\partial L}{\partial N^{k}}=: \mu _k,\qquad \dot\Sigma _k= \dot S_k -\sum_{a=1}^{A_k}\mathcal{J}^{a,k}, \quad k=1,...,K,
\]
where we defined the temperature $T^k$ and the chemical Potential $\mu^k$ of the $k$-th compartment.

From \eqref{EvolEqn_Open}, 
it follows that the internal entropy production $\dot{\Sigma}=\sum_{k=1}^{K}\dot{\Sigma}_{k}$ of the system is given by
\begin{equation}\label{internal_entropy_production}
\begin{aligned}
&I:=\dot{\Sigma}=-\sum_{k=1}^{K}\frac{1}{T^{k}}\left\langle F^{\rm fr}_{k}, \dot q \right\rangle+
\sum_{k<l}^{K}J_{S}^{kl}(T^{l}-T^{k})\left(\frac{1}{T^{l}}-\frac{1}{T^{k}} \right)
+\sum_{k<l}^{K}\mathcal{J}^{k \rightarrow l}\left(\frac{\mu_{k}}{T^{k}}-\frac{\mu_{l}}{T^{l}} \right)\\
&=- \sum_{k=1}^{K}\frac{1}{T^{k}}\sum_{a=1}^{A_k}\Big[\mathcal{J}^{a, k}(\mu^ k- \mu^{a,k})+\mathcal{J}^{a, k}_S(T^k- T^{a,k})\Big]- \sum_{k=1}^{K}\frac{1}{T^{k}}\sum_{b=1}^{B_k}\mathcal{J}^{b, k}_S(T^k- T_{H}^{b,k}).
\end{aligned}
\end{equation}
With this expression of $\dot\Sigma$, the balance of total entropy $S=\sum_{k=1}^{K} S_{k}$ is
\[
\dot{S}=\dot \Sigma+\sum_{k=1}^{K}\sum_{a=1}^{A_k}\mathcal{J}^{a,k}_S+\sum_{k=1}^{K}\sum_{b=1}^{B_k}\mathcal{J}^{b,k}_S.
\]

The total energy associated to the Lagrangian $L(q, \dot q, S_1,...,S_{K}, N^1,...,N^{K})$ is defined as before as
\[
E(q, \dot q, S_1,...,S_{K}, N^1,...,N^{K}):= \left\langle \frac{\partial L}{\partial \dot q}, \dot q \right\rangle -L,
\]
and satisfies the energy balance
\begin{equation}\label{first_law_general}
\begin{aligned}
\frac{d}{dt}E&=\left< F^{\rm ext}, \dot{q} \right>+\sum_{k=1}^{K} \sum_{b=1}^{B_k} \mathcal{J} ^{b,k}_ST_{H}^{b,k}+\sum_{k=1}^{K} \sum_{a=1}^{A_k}\Big[
\mathcal{J} ^{a,k}\mu^{a,k}+ \mathcal{J} ^{a,k}_ST^{a,k}\Big]\\
&=P^{\rm ext}_W+P^{\rm ext}_H+P^{\rm ext}_M,
\end{aligned}
\end{equation}
where $P^{\rm ext}_{W}=\left< F^{\rm ext}, \dot{q} \right>$ is the power exchange due to mechanical work, $P^{\rm ext}_{H}=\sum_{k=1}^{K} \sum_{b=1}^{B_k} \mathcal{J} ^{b,k}_ST_{H}^{b,k}$ is the power exchange due to the heat sources, and $P^{\rm ext}_{M}=\sum_{a=1}^{A_k}\Big[
\mathcal{J} ^{a,k}\mu^{a,k}+ \mathcal{J} ^{a,k}_ST^{a,k}\Big]$  is the power exchange due to matter flowing inside or out of the system through the ports. The energy balance \eqref{first_law_general} is the first law of thermodynamics of this system.

From the second law of thermodynamics, the internal entropy production $\dot{\Sigma}$ in \eqref{internal_entropy_production} must be positive and hence suggests, in the linear regime, the phenomenological relations
\begin{equation}\label{pheno_open_system}
(F^{\rm fr}_k)_{i}=-(\lambda_{k})_{ij}\dot q^{j},\qquad\qquad
\begin{bmatrix}
J^{kl}(T^l-T^k)\\[3mm]
\mathcal{J}^{k \rightarrow l}
\end{bmatrix}= -\mathcal{L}^{kl}\begin{bmatrix}
\frac{1}{T^{l}}-\frac{1}{T^{k}}\\[3mm]
\frac{\mu_{k}}{T^{k}}-\frac{\mu_{l}}{T^{l}}
\end{bmatrix},
\end{equation}
where the symmetric part of the $n\times n$ matrices $\lambda_{k}$ and of the $2\times 2$ matrices $\mathcal{L}^{kl}$ are positive.
The entries of these matrices are phenomenological coefficients determined experimentally, which may in general depend on the state variables. From Onsager's relation, the $2\times 2$ matrix
\[
\mathcal{L}^{kl}=\begin{bmatrix}
\mathcal{L}^{kl}_{HH} & \mathcal{L}^{kl}_{HM}\\[3mm]
\mathcal{L}^{kl}_{MH}& \mathcal{L}^{kl}_{MM}
\end{bmatrix}
\]
is symmetric, for all $k,l$. The matrix elements $\mathcal{L}^{kl}_{HH} $ and $\mathcal{L}^{kl}_{MM}$ are related to the processes of heat conduction and diffusion between the $k$-th and $l$-th compartments. The coefficient $\mathcal{L}^{kl}_{MH}$ and $\mathcal{L}^{kl}_{HM}$ describe the cross-effects, and hence are associated to discrete versions of the process of thermal diffusion and the Dufour effect. Thermal diffusion is the process of matter diffusion due to the temperature difference between the compartments. The Dufour effect is the process of heat transfer due to difference of chemical potentials between the compartments.

\subsection{Examples of Non-Simple Systems with Heat and Matter Transfer}

Here we consider two examples of open non-simple systems experiencing heat conduction, diffusion, and their cross-effects. While the first example concerns the simple case of two compartment, the second describes the case of a composite membrane.

\begin{figure}[h!]
\begin{center}
\includegraphics[scale=0.7]{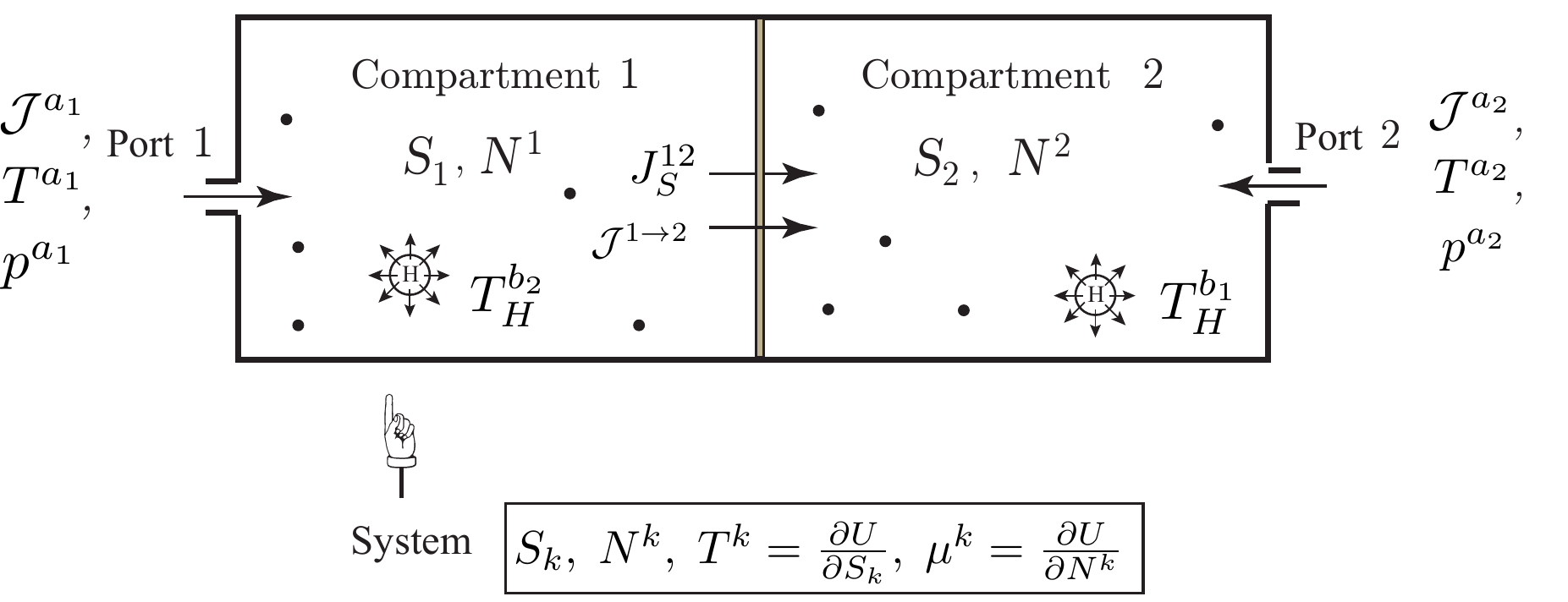}
\vspace{-12pt}
\caption{Non-simple open system with a single chemical component, experiencing diffusion and heat conduction between two compartments, and transfer into or out of the system through two ports.}
\label{MassHeatTransSim}
\end{center}
\end{figure}

\begin{example}[{Heat Conduction and Diffusion between Two Compartments}]{\rm We consider the open system consisting of two compartments as illustrated in Figure~\ref{MassHeatTransSim}. The compartments are separated by a permeable wall through which heat conduction and diffusion is possible. They also have ports $a_1$, $a_2$ through which matter can flow into or out of the system. The fluxes, temperatures, and pressures at the ports are $\mathcal{J}^{a_k}$, $T^{a_k}$, $p^{a_k}$, $k=1,2$. Finally, there are heat sources $b_1$, $b_2$, with entropy flow $\mathcal{J}_S^{b_1}$, $\mathcal{J}_S^{b_2}$ and temperatures $T_{H}^{b_1}$, $T_{H}^{b_2}$.

The Lagrangian of this system is

\[
L(S_1,S_2,N^1, N^2)=- U(S_1,N^1) - U(S_2,N^2) ,
\]
where $U(S,N)$ is the internal energy of the chemical species.
In this case, the system \eqref{EvolEqn_Open} specifies to
\begin{equation}\label{SimpleHeatMassTransferEqn}
\left\{ 
\begin{array}{l}
\displaystyle \dot N^1=  \mathcal{J} ^{ 2 \rightarrow 1}+\mathcal{J} ^{ a_1},\\[3mm]
\displaystyle \dot N^2=  \mathcal{J} ^{ 1 \rightarrow 2}+\mathcal{J} ^{ a_2},\\[3mm]
\displaystyle T^{1}\dot S_{1}=-J^{12}(T^{2}-T^{1})-\mathcal{J} ^{2 \rightarrow 1} \mu _1 +\mathcal{J}^{a_1}(\mathsf{H}^{a_1}-T^1\mathsf{S}^{a_1}-\mu^1)+T^1\mathsf{S}^{a_1}\mathcal{J}^{a_1}+T_{H}^{b_1}\mathcal{J}_S^{b_1},\\[3mm]
\displaystyle T^{2}\dot S_{2}=-J^{12}(T^{1}-T^{2})-\mathcal{J} ^{1 \rightarrow 2} \mu _2 +\mathcal{J}^{a_2}(\mathsf{H}^{a_2}-T^2\mathsf{S}^{a_2}-\mu^2)+T^2\mathsf{S}^{a_2}\mathcal{J}^{a_2}+T_{H}^{b_2}\mathcal{J}_S^{b_2},
\end{array}
\right.
\end{equation} 
where
$$
T^{k}=\frac{\partial U}{\partial S_{k}},\qquad \mu_{k}=\frac{\partial U}{\partial N^{k}},\;\;k=1,2,
$$
are the temperatures and chemical potentials of the $k$-th compartments.

In \eqref{SimpleHeatMassTransferEqn}, the first and second equations are the mole balances in each compartment, while the third and fourth equations are the entropy equations in each compartment. Thus it follows that the equation for the total entropy $S=S_1+S_2$ of the system is
\begin{align*}
\dot S& =J^{12}(T^{1}-T^{2})\left(\frac{1}{T^{1}}-\frac{1}{T^{2}}\right)+\mathcal{J}^{1 \rightarrow 2} \left(\frac{\mu_{1}}{T^{1}}-\frac{\mu_{2}}{T^{2}}\right)\\
&\qquad\qquad\qquad\qquad+\sum_{k=1}^2\left(\frac{1}{T^k}\mathcal{J}^{a_k}(\mathsf{H}^{a_k}-T^k\mathsf{S}^{a_k}-\mu^k) +\mathsf{S}^{a_k}\mathcal{J}^{a_k}+\frac{T_{H}^{b_k}}{T^k}\mathcal{J}^{b_k}_S\right).
\end{align*}
The internal entropy production is found from \eqref{internal_entropy_production} as
\begin{equation}\label{Sigma_compartments}
\begin{aligned}
I:=\dot{\Sigma}&=J^{12}(T^{1}-T^{2})\left(\frac{1}{T^{1}}-\frac{1}{T^{2}}\right)+\mathcal{J}^{1 \rightarrow 2} \left(\frac{\mu_{1}}{T^{1}}-\frac{\mu_{2}}{T^{2}}\right) \\
&\qquad +\sum_{k=1}^2\frac{1}{T^k}\mathcal{J}^{a_k}(\mathsf{H}^{a_k}-T^k\mathsf{S}^{a_k}-\mu^k)+\sum_{k=1}^2\frac{1}{T^k} \mathcal{J}_S^{b_k}(T_{H}^{b_k}-T^k)\ge 0,
\end{aligned}
\end{equation}
from which, in the linear regime, the phenomenological relations are obtained as in \eqref{pheno_open_system}.

The energy balances for each compartment are
\begin{align*}
\frac{d}{dt}U_1&=-J^{12}(T^2-T^1)+\mathcal{J}^{a_1}\mathsf{H}^{a_1}+T_{H}^{b_1}\mathcal{J}_S^{b_1}=P^{2\rightarrow 1}+P_M^{a_1}+P^{b_1}_H,\\
\frac{d}{dt}U_2&=-J^{12}(T^1-T^2)+\mathcal{J}^{a_2}\mathsf{H}^{a_2}+T_{H}^{b_2}\mathcal{J}_S^{b_2}=P^{1\rightarrow 2}+P_M^{a_2}+P^{b_2}_H,
\end{align*}
where $P^{1\rightarrow 2}=J^{12}(T^2-T^1)$ is the power exchanged between the two compartments, $P_M^{a_k}= \mathcal{J}^{a_k}\mathsf{H}^{a_k}$ is the power exchanged between the $k$-th compartment and the exterior through the $a_k$-th port, and~$P^{b_k}_H=T_{H}^{b_k}\mathcal{J}_S^{b_k}$ is the power exchanged between the $k$-th compartment and the heat source $b_k$.
Note that from the definition of $P_H^{b_k}$, we can rewrite the last term of the entropy production Equation \eqref{Sigma_compartments} as
\[
\sum_{k=1}^2P^{b_k}_H\left(\frac{1}{T^k}-\frac{1}{T_{H}^{b_k}}\right).
\]
Note that the relation $P^{1\rightarrow 2}=J^{12}(T^2-T^1)$ related the flux $J^{12}$ used in the variational formulation, to~the total power exchange $P^{1\rightarrow 2}$ between the compartments, due to heat condition, diffusion, and~their~cross-effects.

The total energy balance reads
\[
\frac{d}{dt}U= P_M^{\rm ext}+P_H^{\rm ext}, \qquad P_M^{\rm ext}=\sum_{k=1}^2 P_M^{a_k},\quad P_H^{\rm ext}=\sum_{k=1}^2P^{b_k}_H.
\]}
\end{example}

\begin{figure}[h!]
\begin{center}\includegraphics[scale=0.7]{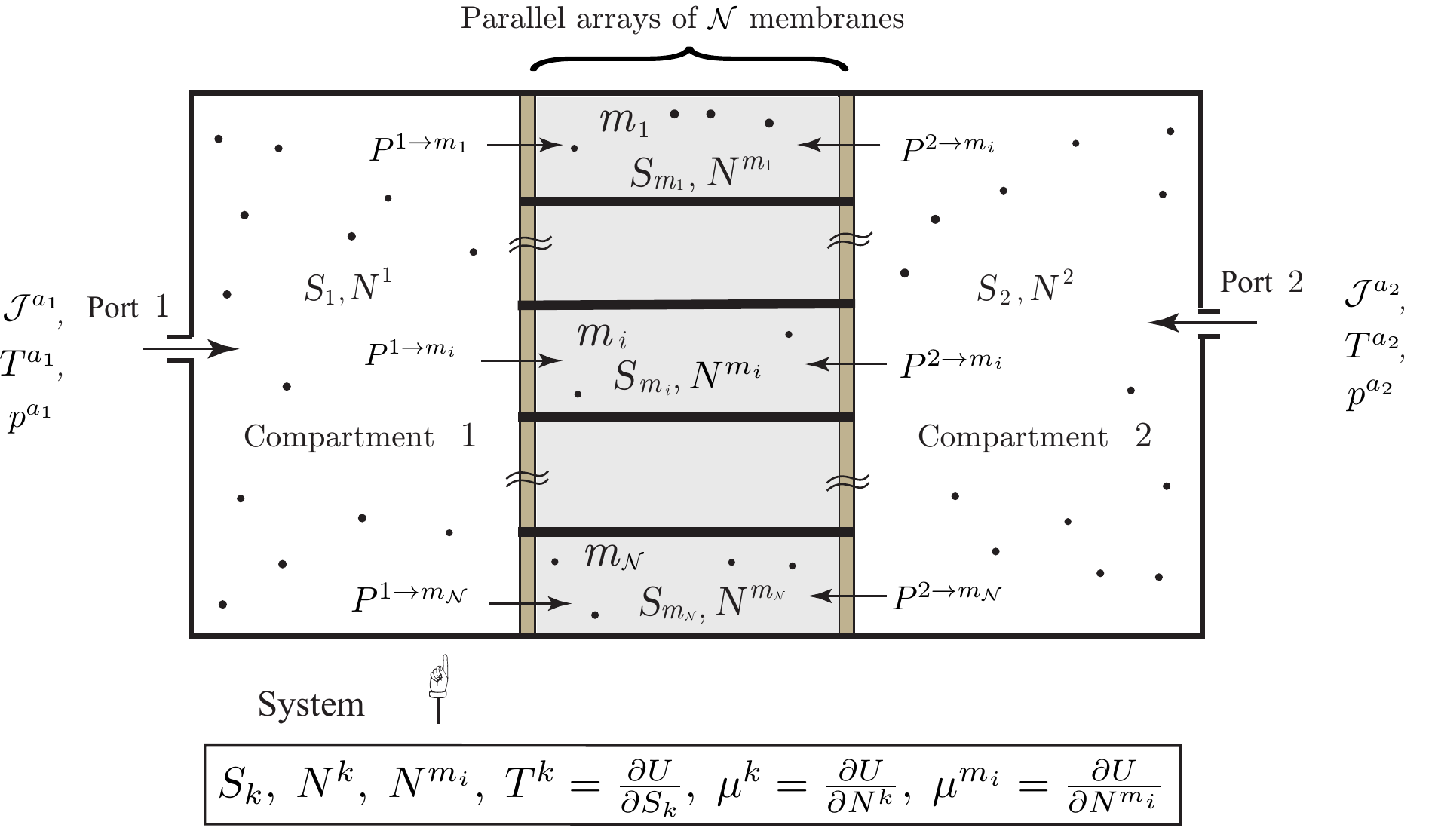}
\caption{Non-simple open system with heat conduction and diffusion of a single chemical species through a composite membrane made of a parallel array {of} $\mathcal{N}$ elements.}
\label{par_membrane_mass_heat_transfer}
\end{center}
\end{figure}

\begin{example}[Heat Conduction and Diffusion through a Composite Membrane]{\rm We consider the open system illustrated in Figure~\ref{par_membrane_mass_heat_transfer}, which describes the heat conduction and diffusion of a single species through a composite membrane made from a  parallel array of $\mathcal{N}$ elements. We also assume that each compartment as a port. For simplicity, we do not consider external heating. It can be included in a similar way with the previous example.

In this case, the system \eqref{EvolEqn_Open} specifies to
\[
\left\{ 
\begin{array}{l}
\displaystyle \dot N^1=  \sum_{i=1}^{\mathcal{N}}\mathcal{J} ^{ m_{i} \rightarrow 1}+\mathcal{J}^{a_1},\qquad \dot N^2=  \sum_{i=1}^{\mathcal{N}} \mathcal{J}^{m_{i} \rightarrow 2}+\mathcal{J}^{a_2},\\[5mm]
\displaystyle \dot N^{m_{i}}=  \mathcal{J}^{1 \rightarrow m_{i}}+ \mathcal{J}^{2 \rightarrow m_{i}},\quad i=1,...,\mathcal{N},\\[3mm]
\displaystyle T^{1}\dot S_{1}=-\sum_{i=1}^{\mathcal{N}}  J^{1m_{i}}(T^{m_{i}}-T^{1})-\sum_{i=1}^{\mathcal{N}} \mathcal{J} ^{m_{i} \rightarrow 1} \mu _1 +\mathcal{J}^{a_1}(\mathsf{H}^{a_1}-T^1\mathsf{S}^{a_1}-\mu^1)+T^1\mathsf{S}^{a_1}\mathcal{J}^{a_1},\\[4mm]
\displaystyle T^{2}\dot S_{2}=-\sum_{i=1}^{\mathcal{N}}  J^{2m_{i}}(T^{m_{i}}-T^{2})-\sum_{i=1}^{\mathcal{N}} \mathcal{J} ^{m_{i} \rightarrow 2} \mu _2 +\mathcal{J}^{a_2}(\mathsf{H}^{a_2}-T^2\mathsf{S}^{a_2}-\mu^2)+T^2\mathsf{S}^{a_2}\mathcal{J}^{a_2},\\[5mm]
\displaystyle T^{m_{i}}\dot S_{m_{i}}=-J^{1m_{i}}(T^{1}-T^{m_{i}})-J^{2m_{i}}(T^{2}-T^{m_{i}})
-\mathcal{J} ^{1 \rightarrow m_{i}} \mu _{m_{i}}-\mathcal{J} ^{2 \rightarrow m_{i}} \mu _{m_{i}},\quad i=1,...,\mathcal{N}.
\end{array}
\right.
\]
The internal entropy production is found from \eqref{internal_entropy_production} as
\begin{equation}\label{entropy_equation_last}
\begin{aligned}
I:=\dot \Sigma&=\sum_{k=1}^2\sum_{i=1}^\mathcal{N} J^{m_ik}(T^{m_i}-T^{k})\left(\frac{1}{T^{m_i}}-\frac{1}{T^{k}}\right)+\sum_{k=1}^2\sum_{i=1}^\mathcal{N}\mathcal{J}^{k \rightarrow m_i} \left(\frac{\mu_{k}}{T^{k}}-\frac{\mu_{m_i}}{T^{m_i}}\right)\\
&\qquad\qquad\qquad\qquad 
+\sum_{k=1}^2 \frac{1}{T^k}\mathcal{J}^{a_k}(\mathsf{H}^{a_k}-T^k\mathsf{S}^{a_k}-\mu^k) ,
\end{aligned}
\end{equation}
from which the entropy equation follows as
\[
\dot S=\dot \Sigma+ \sum_{k=1}^2\mathsf{S}^{a_k}\mathcal{J}^{a_k}.
\]}
\end{example}

\unskip
\vspace{-12pt}

\section{Conclusions}\label{Section_5}

In this paper, we have developed a Lagrangian variational formulation of nonequilibrium thermodynamics for discrete open systems. We have introduced an abstract variational formulation for systems with time-dependent nonlinear nonholonomic constraints, based on a variational constraint and a kinematic constraint, related in a specific way. We have first shown how the proposed variational formulation applies to {simple} open systems, i.e., system well described by a single entropy variable. We have considered simple systems experiencing mechanical interactions, internal diffusion process between several compartments, as well as matter transfer with the exterior through several ports. We have illustrated the proposed framework with several examples such as diffusion through serial and parallel composite membranes. Then, we have extended our variational formulation to the case of {non-simple} open systems, experiencing mechanical interactions, internal diffusion, internal heat transfer, and their cross-effects, which are the discrete versions of thermal diffusion and of the Dufour effect. In each case, our approach yields a systematic and efficient way to derive the complete evolution equations for the open discrete system, independently on its complexity. Such an approach may be especially useful for the treatment of nonequilibrium thermodynamic of biophysical systems.

The approach presented in this paper provides the foundational step for further developments on the nonequilibrium thermodynamics of open systems. As our future works, we project to consider (i) the associated Dirac dynamical system formulation for discrete open systems, see~\cite{GBYo2018a} for closed systems; (ii) the variational numerical discretization for discrete open systems, see~\cite{GBYo2018b} for closed system; (iii) the treatment of continuum open systems.
\vspace{6pt}

\paragraph{Acknowledgments.} Fran\c{c}ois Gay-Balmaz is partially supported by the ANR project GEOMFLUID, ANR-14-CE23-0002-01; \mbox{Hiroaki Yoshimura~is partially} supported by JSPS Grant-in-Aid for Scientific Research (26400408, 16KT0024, 17H01097), the~MEXT ``Top Global University Project'' and Waseda University (SR~2017K-167).

\end{document}